\documentclass[margin= 1in]{interact}

\usepackage{epstopdf}
\usepackage[caption=false]{subfig}

\usepackage[longnamesfirst,sort]{natbib}
\bibpunct[, ]{(}{)}{;}{a}{,}{,}
\renewcommand\bibfont{\fontsize{10}{12}\selectfont}

\theoremstyle{plain}
\newtheorem{theorem}{Theorem}[section]

\theoremstyle{definition}

\theoremstyle{remark}

\usepackage{titlesec}
\usepackage[colorlinks = true,urlcolor  = green,
            citecolor = blue]{hyperref}
\usepackage{siunitx}
\usepackage{url}
\usepackage{xcolor}
\usepackage{float}
\usepackage{setspace}
\usepackage{natbib}
\bibpunct{(}{)}{;}{a}{}{,}
\usepackage{amsmath, amssymb}
\usepackage{graphicx}
\usepackage{caption}
\usepackage{caption}
\usepackage{caption}
\usepackage[normalem]{ulem}
\usepackage{mathbbol}
\usepackage{bm} 

\usepackage{booktabs}
\def\*#1{\mathbf{#1}}
\def\^#1{\amsmathbb{#1}}
\def\##1{\mathbb{#1}}
\DeclareSymbolFontAlphabet{\amsmathbb}{AMSb}%
\usepackage{setspace}
\begin{document}

\articletype{Original Paper}

\title{A Score Based Test for Functional Linear Concurrent Regression}

\author{
\name{Rahul Ghosal\textsuperscript{a}\thanks{CONTACT Rahul Ghosal. Email: rghosal@ncsu.edu} and Arnab Maity\textsuperscript{a}}
\affil{\textsuperscript{a} Department of Statistics, North Carolina State University, Raleigh, NC, USA}
}

\maketitle

\begin{abstract}
We propose a novel method for testing the null hypothesis of no effect of a covariate on the response in the context of functional linear concurrent regression. We establish an equivalent random effects formulation of our functional regression model under which our testing problem reduces to testing for zero variance component for random effects. For this purpose, we use a one-sided score test approach, which is an extension of the classical score test. We provide theoretical justification as to why our testing procedure has the right levels (asymptotically) under null using standard assumptions. Using numerical simulations, we show that our testing method has the desired type I error rate and gives higher power compared to a bootstrapped F test currently existing in the literature. Our model and testing procedure are shown to give good performances even when the data is sparsely observed, and the covariate is contaminated with noise. Applications of the proposed testing method are demonstrated on gait study and a dietary calcium absorption data.

\end{abstract}


\doublespacing
\section{Introduction}
\label{sec:intro}

Functional linear concurrent regression model arises when the response and covariates are both functions of time (or any continuous index), and the value of the response at a particular time point is modeled as a linear combination of the covariates at that specific time point, where the coefficients of the functional covariates are functions of time \citep{Ramsay05functionaldata}. One can view the functional linear concurrent regression model as a series of linear regression for each time point, with the assumption that the coefficient functions are smooth over time.  Multiple methods exist in literature for estimation of these regression coefficient functions in functional linear concurrent  regression and the closely related varying coefficient model \citep{hastie1993varying} using basis functions with roughness penalty \citep{Ramsay05functionaldata}, polynomial spline \citep{huang2002varying, huang2004polynomial}, local polynomial smoothing \citep{wu1998asymptotic, cai2000functional, fan2000simultaneous}, Bayesian modeling \citep{gelfand2003spatial}, covariance representation techniques \citep{csenturk2011varying}, among many others. While estimation of the regression functions is an important problem, in many cases the primary interest might be finding out whether a specific covariate is truly significant or not, i.e., to test for association between a predictor of interest and the response. For example, in the gait study data \citep{Ramsay05functionaldata}, where there are longitudinal measurements of hip and knee angles taken on 39 children, the main purpose of the study is to understand how the joints in hip and knee interact during a gait cycle \citep{theologis2009children}. One natural question to ask here would be, whether the knee angles (response) are at all associated with the hip angles (covariate). Simply building a point-wise confidence interval of the estimated regression function does not answer the question of the overall significance of the covariate. Thus there is a need for developing testing methods to find out significant predictors in this setting.



Formally, our primary goal is to test the null hypothesis that the coefficient function corresponding to a predictor of interest is identically zero, versus the alternative hypothesis that the coefficient function is non-zero for some time point. Literature relating to such global testing in functional concurrent linear regression or closely related varying coefficient model can be traced back to \cite{huang2002varying}, \cite{guo2002functional} among many others. \cite{huang2002varying} employed a resampling subject bootstrap method on an F-type statistic, whereas \cite{guo2002functional} used the connection between linear mixed effects models and smoothing splines, and subsequently used a generalized maximum likelihood ratio test to test for significance of predictors. \cite{kim2016general} extended the bootstrap based test \citep{huang2002varying} to general nonlinear functional concurrent model. Both of these tests rely on a subject-level bootstrap method to obtain the p-values making the test computationally intensive. Recently, \cite{wang2018unified} developed a method for pointwise as well as global testing using empirical likelihood ratio tests. Their method, which is extremely general, uses a wild bootstrap procedure to perform the test for both dense and sparse functional data.
Besides these global testing procedures one can also use confidence bands based methods e.g., \cite{fan2000simultaneous} for building simultaneous confidence bands for the underlying coefficient functions.



In this article, our goal is to build a classical likelihood based testing method for testing of the global effect of covariates, which is also computationally cheap. We model the unknown regression functions using B-spline basis functions and derive an equivalent random effects model. Under such a framework, we show that our testing problem reduces to testing for zero variance components for a set of random effects. There are multiple existing methods in the literature for testing for zero variance components. \cite{crainiceanu2004likelihood}, \cite{greven2008restricted}, and  \cite{staicu2014likelihood} considered testing for variance components using likelihood ratio test (LRT) and restricted likelihood ratio test (RLRT). The main challenge of such tests is that the null distribution is different \citep{crainiceanu2004likelihood} from the commonly used $0.5{\chi}^2_{0}$ : $0.5{\chi}^2_{1}$ approximation or such mixtures of two chi-square distributions, which is used in \cite{guo2002functional}. \cite{crainiceanu2004likelihood} showed the large sample chi-square mixture approximations using the usual asymptotic theory \citep{self1987asymptotic, liang1996asymptotic} for null hypothesis on the boundary of the parameter space do not hold  in general as the inherent assumptions are not valid for linear mixed models. In this article, we propose a score based testing method that is computationally efficient. Our procedure is inspired from the work of \cite{molenberghs2007likelihood} which describes an approach of using a one-sided score test in constrained parameter space. The major advantage of working with the score test is, it does not require computations under the alternative. \cite{zhang2008variance} and \cite{lin1997variance} also used such one-sided score tests for variance component testing in generalized linear mixed models and longitudinal data. However, the methods mentioned above assumed that the responses are independent given the random effects and that the variances have some parametric form. In contrast, in our functional regression framework, we assume unknown non-trivial covariance structure and estimate the covariance function nonparametrically. The assumption of non-trivial dependence is crucial in functional data because of complex correlation structures that might be present in real data. We derive the asymptotic distribution of the test statistic under the null hypothesis; we show that the commonly used chi-squared approximation of the score test statistic is not appropriate in our situation. However, the null distribution of our test statistic is easy to simulate from. Thus the calculation of p-value for our testing procedure is computationally efficient. We show that asymptotically our testing procedure has the correct type I error rate. Using numerical simulations, we illustrate that our testing method has the desired type I error rates for finite sample sizes and that our proposed testing procedure has higher power than the bootstrapped F-test of \cite{kim2016general}.


The rest of the article is organized as follows. In Section
\ref{sec:method}, we discuss our model specification, present
our testing method and derive theoretical properties related to
our test statistic. In Section \ref{sec:sim_stud}, we present a
simulation study under various sampling design scenarios and
give the simulation results. In Section \ref{sec:real_dat}, we
demonstrate our proposed test by applying it to the two real
data examples: gait data and calcium absorption study and
summarize our findings. We conclude by a discussion about some
limitations and some possible extensions of our work in Section
\ref{sec disc}.
\section{Methodology}
\label{sec:method}
\subsection{Modeling framework}
\label{subsec:M_frame}
Suppose that the observed data for the $i$th subject,
$i\hspace*{- 4 mm}=\hspace*{- 4 mm}1, \ldots, n$, is $\{Y_i(t),
X_{i1}(t),\ldots,X_{ip}(t)\}$, \hspace*{- 1 mm}where $Y(\cdot)$
is a functional response and
$X_1(\cdot)$,$\ldots,X_p(\cdot)$ are the corresponding functional covariates. In practice, the functions for the $i$th subject are observed only on a finite set of points $t_{ij}$, $j = 1, \ldots, m_i$. We assume that $t_{ij} \in {\mathcal T}$, a bounded and closed set. For the rest of the article, we assume $\mathcal{T} = [0,1]$ without loss of generality. To start with, we will assume that $t_{ij} = t_j$, and that the covariates $X_{ik}(\cdot)$ are measured without error. We discuss the cases when the functions are observed on irregularly spaced grid, and with additional measurement errors in Section \ref{subsec:ext}. We consider a linear functional concurrent regression model,
\begin{equation*}
Y_i(t)=\beta_{0}(t)+\sum_{k=1}^{p}X_{ik}(t)\beta_{k}(t)+ \epsilon_{i}(t),
\end{equation*}
where $\beta_{k}(t)$, $k=0,1,2,\ldots,p$, are smooth functions representing functional intercept and functional slope parameters, respectively. We assume $X_{ik}(\cdot)$ are independent and identically distributed (i.i.d.) copies of $\mathcal{X}_k(\cdot)$, $k=1,2,\ldots,p$, where $\mathcal{X}_{k}(\cdot)$ is a stochastic processes with finite second moment. For simplicity we illustrate our testing method for the single covariate model
\begin{equation}
Y_i(t)=\beta_{0}(t)+X_{i}(t)\beta_{1}(t)+ \epsilon_{i}(t),
\end{equation}
which can be easily extended to the multiple covariate situation above and this is discussed in Section \ref{subsec:multcov}. We further assume $\epsilon_{i}(\cdot)$ are i.i.d. copies of $\epsilon(\cdot)$, which is a mean zero Gaussian process plus some Gaussian white noise, that is,
$\epsilon(t)= V(t) + w_t,$ where $V(\cdot) \sim \mathcal{N}(0,G(\cdot,\cdot))$
and $w_t$ are i.i.d. $\mathcal{N}(0,\sigma^2)$ random errors. Thus the covariance function of the error process is given by $\Sigma(s,t)=cov(\epsilon(s),\epsilon(t))=G(s,t) + \sigma^2 I(s=t)$. Our primary interest lies in testing, $$H_{0} : \beta_{1}(t) =0 \hbox{ for all $t$ \;\; versus \;\; } H_{1} : \beta_{1}(t)\neq 0 \hbox{ for some $t$}.$$
In general testing $H_{0}$ is difficult since $\beta_1(t)$ is an
infinite dimensional parameter. In this article, we show that by
modeling the coefficient function with splines and using a random
effects model, the testing problem can be reduced to a variance
component test. This reduction in parameter dimension not only helps
in getting satisfactory performance of our testing method but also
is computationally cheaper than doing the bootstrapped F test
existing in literature. Subsequently we develop a one sided score
test for testing our null hypothesis. 
\vspace*{-3.5 mm}
\subsection{Equivalent random effects model}
An usual method \citep{Ramsay05functionaldata} to estimate $\beta_{0}(t)$ and $\beta_{1}(t)$ in model (1) is by minimizing the penalized residual sum of squares, $$\sum_{i=1}^{n}||Y_i(\cdot)-\beta_{0}(\cdot)-X_{i}(\cdot)\beta_{1}(\cdot)||_{F_2}^{2}+\lambda_{0} \int \{\beta_{0}^r(t)\}^{2}dt +\lambda_{1}\int \{\beta_{1}^r(t)\}^{2}dt,$$ where $\lambda_{0}$, $\lambda_{1}$ are unknown penalty parameters penalizing $r$-th derivative of the coefficient functions and $||\cdot||_{F_2}$ denotes the functional $L_2$ norm. Suppose for $\ell=0,1$ $\{B_{k\ell}(t),  k=1,2,\ldots,k_\ell\}$ is a set of known basis functions. We approximate the unknown coefficient functions using finite basis function expansion as $\beta_\ell(t)= \sum_{k=1}^{K_{\ell}} b_{k\ell}B_{k\ell}(t)=\*B_{\ell}^T(t)\*b_{\ell},\hspace{2mm} \ell = 0, 1$, where $B_{\ell}(t)=[B_{1\ell}(t),B_{2\ell}(t),\ldots ,B_{K_\ell\ell }(t)]^T$ and $b_{\ell}=(b_{1\ell},b_{2\ell},\ldots ,b_{k_\ell \ell})^T$ is a vector of unknown coefficients. In this article, we use cubic B-spline basis functions, however, other basis functions can be used as well.
Thus we can write $X_{i}(t)\beta_{1}(t)=\sum_{k=1}^{k_{1}} b_{k1}X_{i}(t)B_{k1}(t)=\*{X_{i}^{*}}^T(t)\*b_{1}$, where $\*X_{i}^{*}(t)= [X_{i}(t)B_{11}(t),X_{i}(t)B_{21}(t), \ldots, X_{i}(t)B_{k_{1}1}(t)]^T$. We can then rewrite our model (1) as 
$Y_i(t)=\*B_{0}^T(t)\*b_{0}+\*{X_{i}^{*}}^T(t)\*b_{1}+\epsilon_{i}(t)$. The unknown basis coefficients can then be estimated by minimizing the  penalized error sum of squares $\sum_{i=1}^{n}||Y_i(\cdot)-\*B_{0}^T(\cdot)\*b_{0}-{\*X_{i}^{*}}^T(\cdot)\*b_{1}||_{2}^{2}+\lambda_{0}
\*b_{0}^{T}\^P_{0}\*b_{0}+\lambda_{1}\*b_{1}^{T}\^P_{1}\*b_{1},$
where $\^P_{0}$ and $\^P_{1}$ are the penalty matrices coming from penalizing the $r$-th derivative of the functions $\beta_{0}(t)$ and $\beta_{1}(t)$. In particular $\^P_{\ell}=\int \*B^{r}_{\ell}(t)\*B^{r}_{\ell}(t)^{T}dt $ and thus
$\int (\beta_{\ell}^r(t))^{2}dt = \*b_{\ell}^{T}\^P_{\ell}\*b_{\ell}$.
 Since we only observe data on a fine regular grid $S= \{t_{1},
t_{2},\ldots,t_{m} \}$ in practice, the minimization is carried out by minimizing 
$$\sum_{i=1}^{n}\sum_{j=1}^{m}\{Y_i(t_{j})-\*B_{0}^T(t_{j})\*b_{0}-\*{X_{i}^{*}}
^T(t_{j})\*b_{1}\}^{2}+
\lambda_{0}
\*b_{0}^{T}\^P_{0}\*b_{0}+\lambda_{1}\*b_{1}^{T}\^P_{1}\*b_{1}.$$
Define $ \*Y_i=[Y_i(t_{1}),Y_i(t_{2}),\ldots, Y_i(t_{m})]^{T}$, $\^B_{0}=[\*B_{0}(t_{1})| \*B_{0}(t_{2})|\ldots|\*B_{0}(t_{m})]^{T}$,\\
$\^X_{i}=[\*X_{i}^{*}(t_{1})|\*X_{i}^{*}(t_{2})|\ldots|\*X_{i}^{*}(t_{m})]^{T}$. Thus the least square criterion becomes
$$\sum_{i=1}^{n}||\*Y_i-\^B_{0}\*b_{0}-\^X_{i}\*b_{1}||_{2}^{2}+
\lambda_{0}
\*b_{0}^{T}\^P_{0}\*b_{0}+\lambda_{1}\*b_{1}^{T}\^P_{1}\*b_{1}.$$
Now since the matrices $\^P_{0}$, $\^P_{1}$ are singular (for $r\geq 1$), the equivalent random effects model corresponding to this minimization problem would be rank deficient. As our primary interest lies in testing, we propose to penalize the coefficient functions directly, namely we use $r=0$ and ridge penalty, consequently $\^P_{\ell}=\int \*B_{\ell}(t)\*B_{\ell}(t)^{T}dt$. Our simulations show we are able to maintain correct type I error rates of the proposed testing method using this strategy. It then follows that the normal equations are identical to those from the equivalent random effects model $\*Y_i=\^B_{0}\*b_{0}+\^X_{i}\*b_{1}+\bm\zeta_i$,
 where $\bm\zeta_i\sim \mathcal{N}_{m}(0,\sigma_\epsilon^{2}\^I_m)$,  $\*b_{0}\sim\mathcal{N}_{k_0}(0,\sigma_0^{2}\mathbb{\Sigma}_0)$,
$\*b_{1}\sim\mathcal{N}_{k_1}(0,\sigma_1^{2}\mathbb{\Sigma_1})$ and $\*b_{0}$, $\*b_{1}$, $\zeta_i$ are independent. We denote $\#\Sigma_0=\^P_{0}^{-1}$, $\#\Sigma_1=\^P_{1}^{-1}$, $\sigma_0^{2}=\frac{\sigma_\epsilon^{2}}{\lambda_{0}}$, $\sigma_1^{2}=\frac{\sigma_\epsilon^{2}}{\lambda_{1}}$. 
Using Cholesky decomposition of $\#\Sigma_0$, $\#\Sigma_1$ and 
appropriately reparameterizing (${\^C_{0}}={\^B_{0}}\#\Sigma_0^{1/2}$, $\^Z_{i}=\^X_i\#\Sigma_1^{1/2}$, $\bm{\gamma}_{k}=\#\Sigma_k^{-1/2}\*b_k$), the model can be rewritten as
$
\*Y_i=\^C_{0}\bm{\gamma}_{0}+\^Z_{i}\bm{\gamma}_{1}+{\bm\zeta_i},
$
where $\bm\zeta_i\sim \mathcal{N}_{m}(0,\sigma_\epsilon^{2}\^I_m)$, $\bm\gamma_{0}\sim \mathcal{N}_{k_0}(0,\sigma_0^{2}\^I_{k_0})$, $\bm\gamma_{1}\sim \mathcal{N}_{k_1}(0,\sigma_1^{2}\^I_{k_1})$, and all the random effects are independent. Thus our test $H_0$ can be carried out via testing of a single variance component, namely testing $H_{0} :\sigma_1^{2} =0 $ against the alternative $H_{1} : \sigma_1^{2}> 0 $.

Now for our testing problem the errors are not independent and more likely to be temporally correlated. Therefore to get the correct likelihood we need to use the true covariance kernel $\Sigma(s,t)$ which takes into account this temporal dependence of the residual vector $\bm\epsilon_i$s. This motivates us to use the following random effects model 
\begin{equation}
    \*Y_i=\^C_{0}\bm{\gamma}_{0}+\^Z_{i}\bm{\gamma}_{1}+{\bm\epsilon_i},
\end{equation}
where $\bm\epsilon_i\sim \mathcal{N}_{m}(0,\#\Sigma_{m\times m})$, $\bm\gamma_{0}\sim \mathcal{N}_{k_0}(0,\sigma_0^{2}\^I_{k_0})$, $\bm\gamma_{1}\sim \mathcal{N}_{k_1}(0,\sigma_1^{2}\^I_{k_1})$ and all of them are independent. Here $\#\Sigma_{m\times m}$ denotes the covariance kernel $\Sigma(s,t)$ evaluated at $S= \{t_{1},
t_{2},\ldots, t_{m} \} $. For the moment let us assume $\#\Sigma_{m\times m}$ to be known. Of course in reality $\#\Sigma_{m\times m}$ will be unknown and we will need to estimate it. We illustrate in Section \ref{subsec:est_cov} how to estimate $\#\Sigma_{m\times m}$ using functional principal component analysis (FPCA). Writing equation (2) in stacked form for $i= 1,2,3,\ldots,n$ we have 
\begin{equation}
\*Y=\^B\bm\gamma_{0}+\^Z\bm\gamma_{1}+\mathbf{\mathcal{E}},
\end{equation}
where  $\^B=[\^C_0^T|\^C_0^T|\ldots|\^C_0^T]^T$, $\^Z=[\^Z_1^T|\^Z_2^T|\ldots|\^Z_n^T ]^T$, $\*Y=(\*Y_1^T,\*Y_2^T,\ldots,\*Y_n^T)^T$ and $\mathbf{\mathcal{E}}=(\bm\epsilon_1^T,\bm\epsilon_2^T,\ldots,\bm\epsilon_n^T)^T$, $\mathbf{\mathcal{E}}\sim \mathcal{N}_{N}(0,\#\Sigma)$
($N=mn$), where $\#\Sigma$=diag \{${\#\Sigma_{m\times m},\#\Sigma_{m\times m},\ldots, \#\Sigma_{m\times m}}$\}. Note that $\^Z=\^X\#\Sigma_1^{1/2}$ and $\^B=\mathbf{\mathcal{B}}\#\Sigma_0^{1/2}$, where $\^X,\mathbf{\mathcal{B}}$ are defined similarly by stacking $\^X_i$ and $\^{B}_0$ s. In this set up, we are interested in testing $H_{0} :\sigma_1^{2} =0 $ against the alternative $H_{1} : \sigma_1^{2}> 0 $, which as demonstrated, is equivalent to testing the null hypothesis of no effect of a covariate on the response under functional linear concurrent model. Next we develop a score based test for conducting the test of variance component.

\subsection{Testing method}
\label{subsec:Test_m}
We develop our testing method treating $\^Z$ as nonrandom (fixed). Namely our test is a conditional test based on observed $\^Z$ \{i.e., observed $X_{i}(t)$\}. We show that our conditional testing method has the right levels under null which in turn ensures the unconditional test would also enjoy this property. Marginally $\*Y\sim\mathcal{N}(0,\^V)$, which follows from equation (3) with $\^V=\^V(\tau_0,\tau_1)=\#\Sigma+\tau_0\^B\^B^{T}
 +\tau_1\^Z\^Z^{T}$, where $\tau_0=\sigma_0^{2}$ and $\tau_1=\sigma_1^{2}$. So the marginal log-likelihood of $\*Y$ (upto a constant) is :
\begin{equation}
L_{ML}(\tau_0,\tau_1)=-1/2(\ln|\^V|+\*Y^{T}\^V^{-1}\*Y).
\end{equation}
Based on this likelihood, we want to test $H_{0} :\tau_1 =0$ vs $H_{1} : \tau_1> 0 $. 

Let $\bm\theta=(\tau_0,\tau_1)^T$ and $\tilde{\bm\theta}$ denote the maximum likelihood estimate (M.L.E) of $\bm\theta$ under $H_0$.
The score function of $\tau_1$ is 
\begin{align*}
S_{\tau_1}(\tau_0,\tau_1)&=-1/2\{tr(\^V^{-1}\^M)
-\*Y^{T}\^V^{-1}\^M\^V^{-1}\*Y\} \\
&=-1/2\{tr(\^ Z^{T}\^V^{-1}\^Z)-
(\^V^{-1/2}\*Y)^{T}\^V^{-1/2}\^Z\^Z^{T}\^V^{-1/2}(\^V^{-1/2}\*Y)\},
\end{align*}
where $\^M=\^Z\^Z^{T}$.
The information matrix $\^I(\bm\theta)$ corresponding to the likelihood in (4) is partitioned as,
\[
 \^I(\bm\theta)=  \begin{Bmatrix} 
      I_{11}(\bm\theta) &  I_{12}(\bm\theta)  \\ 
      I_{21}(\bm\theta)^{T} & I_{22}(\bm\theta)  
      \end{Bmatrix},
\]
where $I_{11}(\bm\theta)=tr\{ (\^B^{T}\^V^{-1}\^B)^2\}/2, 
I_{22}(\bm\theta)=tr\{ (\^Z^{T}\^V^{-1}\^Z)^2\}/2$,  
 $I_{21}(\bm\theta)=I_{12}(\bm\theta)\\=tr\{ (\^B^{T}\^V^{-1}\^Z)(\^B^{T}\^V^{-1}\^Z)^T\}/2$. 
Then the classical score test statistic is given by,
$$T_S=\frac{S_{\tau_1}^2(\tilde{\bm\theta})}{I_{22}(\tilde{\bm\theta})-I_{21}(\tilde{\bm\theta})^T I_{11}^{-1}(\tilde{\bm\theta})I_{12}(\tilde{\bm\theta})}.$$ 

 As the parameter space is constrained and the null hypothesis is on the boundary of the parameter space, following \cite{molenberghs2007likelihood} we define our one sided score test statistic as  
\begin{equation}
T_S=
\begin{cases}
\frac{S_{\tau_1}^2(\tilde{\bm\theta})}{\Lambda(\tilde{\bm\theta})}\hspace{3.6 cm} \text{if $S_{\tau_1}(\tilde{\bm\theta})\geq0$}\\
0 \hspace{4.4 cm} \text{if $S_{\tau_1}(\tilde{\bm\theta})<0$},\\
\end{cases}
\end{equation}
where $\Lambda(\tilde{\bm\theta}) = I_{22}(\tilde{\bm\theta})-I_{21}(\tilde{\bm\theta})^TI_{11}^{-1}(\tilde{\bm\theta})I_{12}(\tilde{\bm\theta}).$
The reason behind using such one sided score test statistic is if the score is negative at the M.L.E under null, then the value of the score gives no evidence in favour of the alternative and therefore the statistic is set to zero in such cases. An illustration for constrained one parameter case is given in Figure \ref{fig:fig0}.
\renewcommand{\labelenumi}{\alph{enumi})} 
We assume the true covariance matrix $\#\Sigma$ to be known for the time being for establishing asymptotic distribution of our test statistic but in reality it is generally unknown, so we will need to estimate it from data by some consistent estimator $\hat{\#\Sigma}$ and plug that in for $\#\Sigma$ in $T_S$. Next, we posit two theorems which establish asymptotic distribution our test statistic.
\begin{theorem}
Suppose the following conditions are true :
\begin{enumerate}
\item The null hypothesis is true i.e., $H_{0} :\tau_1 =0$ holds and $\bm\theta_0 =(\tau_0^{*},0)$ is the true value of $\bm\theta$,
\item $\#\Sigma$ be the true covariance matrix of the residual vector $\mathbf{\mathcal{E}}$ in equation (3).
\end{enumerate}
Then $T_S(\bm\theta_0,\#\Sigma) = \frac{S_{\tau_1}^{2}({\bm\theta_0})}{\Lambda({\bm\theta_0})}I(S_{\tau_1}({\bm\theta_0})\geq0) \overset{d}{=}   (1/2)^2\frac{\left(\sum_{\ell=1}^{k_{1}}\lambda_{\ell}x_{\ell}^{2}-\sum_{\ell=1}^{k_{1}}\lambda_{\ell}\right)^2}{\Lambda_n(\bm\theta_0)}I(\sum_{\ell=1}^{k_{1}}\lambda_{\ell}x_{\ell}^{2}\geq \sum_{\ell=1}^{k_{1}}\lambda_{\ell})$,\newline
where $x_{\ell}\stackrel{iid}{\sim}\mathcal{N}(0,1)$ and $\lambda_\ell$ are eigenvalues of $\^Z^{T}\^V(\theta_0,\#\Sigma)^{-1}\^Z/n$ and 
$\Lambda_n(\bm\theta_0)=\Lambda(\bm\theta_0)/n^2
=\frac{1}{2}tr\{(\^Z^{T}
\^V^{-1}\^Z/n)^2\}$\hspace{1mm}-\hspace{1mm} $\frac{\left[\frac{1}{2}tr\{(\^B^{T}\^V^{-1}\^Z/n)(\^B^{T}\^V^{-1}\^Z/n)^T\}\right]^2}{\frac{1}{2}tr\{ (\^B^{T}\^V^{-1}\^B/n)^2\}}$ 
\end{theorem}

The proof of the Theorem 2.1 is given in Appendix \ref{app}.
\begin{theorem}
Suppose the following conditions are true :
\begin{enumerate}
\item The null hypothesis is true i.e $H_{0} :\tau_1 =0$ holds and $\bm\theta_0 =(\tau_0^{*},0)$ is the true value of $\bm\theta$,
\item $\tilde{\bm\theta}$ is  $\sqrt[]{n}$ consistent estimator of $\bm\theta_0$ under null and the estimator $\hat{\#\Sigma}$ is a consistent estimator of $\#\Sigma$ in the sense $||\hat{\#\Sigma}^{-1}- \#\Sigma^{-1}||_{2} = o_{p}(1)$ (spectral norm).
\end{enumerate}
Then $T_S(\tilde{\bm\theta},\hat{\#\Sigma})\overset{d}{\rightarrow}T_S(\bm\theta_0,\#\Sigma)$.
\end{theorem}

The proof is mainly based on application of Slutsky's theorem and matrix norm inequalities. A detailed proof is given in Appendix S1 of supplemental material. As mentioned in Theorem 2.1 the null distribution of the test statistic is given by  $(1/2)^2\frac{\left(\sum_{\ell=1}^{k_{1}}\lambda_\ell x_{\ell}^{2}-\sum_{\ell=1}^{k_{1}}\lambda_\ell\right)^2}{\Lambda_n(\theta_0)}I(\sum_{\ell=1}^{k_{1}}\lambda_{\ell} x_{\ell}^{2}\geq \sum_{\ell=1}^{k_{1}}\lambda_\ell)$. Because $\bm\theta_0$ and $\lambda_\ell$ are unknown in reality, we approximate the null distribution using plug-in estimates 
of $\tilde{\bm\theta}$ and $\hat{\#\Sigma}$, i.e, we use the approximate null distribution 
\begin{equation}
(1/2)^2\frac{\left(\sum_{\ell=1}^{k_{1}}\tilde{\lambda_\ell}x_{\ell}^{2}-\sum_{\ell=1}^{k_{1}}\tilde{\lambda_\ell}\right)^2}{\Lambda_n(\tilde{\bm\theta})}I( \sum_{\ell=1}^{k_{1}}\tilde{\lambda_\ell}x_{\ell}^{2}\geq  \sum_{\ell=1}^{k_{1}}\tilde{\lambda_\ell}),
\end{equation} 
where $\tilde{\lambda_\ell}$ are eigenvalues of $\^Z^{T}\^V(\tilde{\bm\theta},\hat{\#\Sigma})^{-1}\^Z/n$ and $x_{\ell}\stackrel{iid}{\sim}\mathcal{N}(0,1)$ for $\ell=1,2,\ldots, k_1$. This is justified as it can be shown $\tilde{\lambda_\ell}\overset{p}{\rightarrow}\lambda_\ell$ and $\Lambda_n(\tilde{\bm\theta})\overset{p}{\rightarrow}\Lambda_n(\bm\theta_0)$, see the proof in Appendix S1 of supplemental material. Our simulations show that we are able to get the correct type I error rates and good power of our test using the above strategy.

Simulation from the null distribution of our test statistic is easy and computationally efficient, as we only need to calculate $k_1$ eigenvalues of the matrix $\^Z^{T}\^V(\tilde{\bm\theta},\hat{\#\Sigma})^{-1}\^Z/n$, and simulate $x_{\ell}\stackrel{iid}{\sim}\mathcal{N}(0,1)$ for $\ell=1,2,\ldots, k_1$. For calculation of $\^V(\tilde{\bm\theta},\hat{\#\Sigma})^{-1}$, we use the Woodbury matrix identity and also the fact $\hat{\#\Sigma}$ is block diagonal, which greatly speeds up the calculation. It is well known \citep{zhang2003hypothesis}, that the usual asymptotic ${\chi}^2$ distribution of the score test do not work here, so the approximate null distribution in (6) is more appropriate. 

As the test statistic is one sided and in particular not continuous at zero, the p-value under null is asymptotically distributed as mixture distribution of degenerate one and $U(0,\alpha)$, where
$\alpha=P_{H_{0}}\{S_{\tau_1}(\bm\theta_{0})\geq0\}= P(\sum_{\ell=1}^{k_{1}}\lambda_{\ell} x_{\ell}^{2}\geq\sum_{\ell=1}^{k_{1}}\lambda_\ell)$, see Appendix S2 in supplemental material for a detailed proof. As $k_1$ increases it follows by application of CLT, $\alpha\rightarrow \frac{1}{2}$ and the null distribution of our test statistic asymptotically converges to $0.5{\chi}^2_{0}$ : $0.5{\chi}^2_{1}$. In reality choice of $k_1$ will depend on the type of design (dense or sparse) and number of observed time points for each subjects. So such a convergence do not hold \citep{crainiceanu2004likelihood} for subjects observed only on a finite set of points. Therefore, we use the approximate null distribution (6) to perform our test.

\subsection{Estimation of Covariance Matrix}
\label{subsec:est_cov}
In reality $\#\Sigma$ is unknown, and we need a consistent estimator $\hat{\#\Sigma}$. In the context of functional data, we want to estimate $\Sigma(\cdot,\cdot)$ completely non-parametrically. If we had the original residuals $\epsilon_{ij}$ available, we could use functional principal component analysis (FPCA), e.g.,  \cite{yao2005functional} or \cite{zhang2007statistical} to estimate $\Sigma(s,t)$. The error process $\epsilon(t)$ was defined as $\epsilon(t)=V(t) + w_t$. We assume the covariance kernel $G(s,t)$ of the smooth part $V(t)$ is a Mercer kernel \citep{j1909xvi}. Then by Mercer's theorem $G(s,t)$ must have a spectral decomposition
$$G(s,t)=\sum_{k=1}^{\infty}\lambda_k\phi_k(s)\phi_k(t),$$
where $\lambda_1\geq\lambda_2\geq \ldots0$ are the ordered eigenvalues and $\phi_k(\cdot)$s are corresponding eigenfunctions. Thus we have $\Sigma(s,t)=\sum_{k=1}^{\infty}\lambda_k\phi_k(s)\phi_k(t) + \sigma^2 I(s=t)$. Given $\epsilon_{t_{ij}}=V(t_{ij}) + w_{ij}$, one could employ FPCA based methods to get $\hat{\phi}_k(\cdot)$, $\hat{\lambda}_k$s and $\hat{\sigma^2}$. \cite{hall2006properties} established $L^{2}$ convergence of the FPCA estimates, in particular of the eigenfunctions and eigenvalues, in both the sparse and dense functional data setting under appropriate regularity conditions on the sampling design. \cite{li2010uniform} established uniform convergence rates for eigenfunctions and eigenvalues under more general framework where the number of observations for each function can be sampled at any rate relative to the sample size. More specifically \cite{li2010uniform} showed it is possible to get consistent estimators $\hat{\phi}_k(\cdot)$, $\hat{\lambda}_k$ and $\hat{\sigma^2}$ under both sparse and dense functional data settings. So a consistent estimator of 
$\Sigma(s,t)$ can be formed as
$$\hat{\Sigma}(s,t)=\sum_{k=1}^{K}\hat{\lambda_k}\hat{\phi_k}(s)\hat{\phi_k}(t) + \hat{\sigma^2} I(s=t),$$ 
where $K$ is large enough for the convergence to hold and is typically chosen such that percent of variance explained (PVE) by the selected eigencomponents exceeds some pre-specified value such as $99\%$ or $95\%$. In reality we don't have  the original residuals $\epsilon_{ij}$ and use the full model (1) to obtain residuals $e_{ij}=Y_{i}(t_j)-\hat{Y_{i}}(t_j)$. Then treating $e_{ij}$ as our original residuals, we obtain $\hat{\Sigma}(s,t)$ using FPCA. Our simulations show good results using this approach and we are able to maintain the correct levels of the test under both sparse and dense sampling design scenarios.
\subsection{Extension to Sparse and Noisy Covariate}
\label{subsec:ext}
In developing our method we assumed that covariates are measured without noise, and data is observed on a regular dense grid of points $S= \{t_{1},t_{2},\ldots, t_{m} \} \subset \mathcal{T}=[0,1] $. Although in reality data might be observed sparsely, and the covariates may be contaminated with measurement error. Our testing method can be extended to these situations in the following ways.
\subsection*{Case 1: Sparse design, no measurement error}
We assume response $Y_i(t)$ and the covariate $X_{i}(t)$ are observed in $S_i= \{t_{i1},t_{i2},\ldots, t_{im_{i}} \} \subset \mathcal{T}=[0,1]$ for each $i=1,2,\ldots,n$
and $\max_{1 \leq i \leq n}m_{i}\leq M$, for some fixed $M$. In this case the only difference in our model is that ${\*Y_i}$ is a $m_i\times 1$ dimensional vector and that ${\bm\epsilon_i}\sim \mathcal{N}_{m_i}(0,\#\Sigma_{i})$ independently, where $\#\Sigma_{i}$ is the covariance kernel $\Sigma(s,t)$ evaluated at $S_i= \{t_{i1},
t_{i2},\ldots,t_{im_i} \} $. So our model described in (2) still holds with $\#\Sigma$=diag \{${\#\Sigma_{1},\#\Sigma_{2},\ldots,\#\Sigma_{n}}$\}.
As discussed in Section \ref{subsec:est_cov}, if $\Sigma(s,t)$ can be consistently estimated by $\hat{\Sigma}(s,t)$ then $\hat{\#\Sigma_{i}}$ would be a consistent estimator of $\#\Sigma_{i}$  and our testing method is still valid.
\subsection*{Case 2: Dense design with measurement error}
Suppose now data $Y_i(t)$ and covariate $X_{i}(t)$ are observed in  a fine regular grid $S= \{t_{1},t_{2},\ldots,t_{m}\} \subset \mathcal{T}=[0,1]$  and the covariate is observed with measurement error. Namely instead of observing 
$X_{i}(t_j)$ we observe $U_{ij}=X_{i}(t_j)+\delta_{ij}$,
where $\delta_{ij}$ are i.i.d. mean zero random errors with variance $\kappa^2$. There exists several methods to reconstruct the original curve $X_{i}(\cdot)$ from the observed curve with measurement error. \cite{zhang2007statistical} proposed to use local polynomial kernel smoothing technique for individual function reconstructions. They showed that under appropriate conditions and suitable choice of bandwidth, the smoothed trajectories, $\hat{X_i}(\cdot)$  will estimate the true curves ${X_i(\cdot)}$ with negligible error. Thus we can use the reconstructed curves $\hat{X_{i}}(\cdot)$ and the effect of such a substitution is asymptotically negligible.
\subsection*{Case 3 : Sparse design with measurement error}
More generally, we consider the case where functional data is observed in irregular and sparse grid of points and covariate is observed with measurement error. Here we have observed response \{($Y_i(t_{ij}),t_{ij}), j=1,2,\ldots,m_i$\} and observed covariate \{($U(t_{ij}),t_{ij}), j=1,2,\ldots,m_{1i}$\}. Again we assume that $U_{ij}=X_{i}(t_{ij})+\delta_{ij}$, where $\delta_{ij}$ are i.i.d. mean zero random errors with variance $\kappa^2$. In such a sparse sampling design, it is generally assumed \citep{kim2016general} although the individual  number of observations $m_{i}$ is small, $\bigcup_{i=1}^{n}\bigcup_{j=1}^{m_i}{t_{ij}}$ is dense in $\mathcal{T}=[0,1]$. Then reconstructing of the original curves from the observed sparse curves can be done by using  FPCA methods of \cite{yao2005functional} or \cite{hall2006properties}. The methods are based on estimating the mean and covariance functions using local linear smoothing, and subsequently estimating the eigenvalues and eigenfunctions from a spectral decomposition of estimated covariance matrix. 

As mentioned earlier, \cite{li2010uniform} proved uniform convergence of the mean, eigenvalues and eigenfunctions for both dense and sparse design under suitable regularity conditions. For prediction of the scores \cite{yao2005functional} introduced the PACE method which ensures the estimated scores asymptotically goes to BLUP of the original scores. Then these estimates can be put together using  Karhunen-Lo$\grave{e}$ve expansion to get estimates $\hat{X_i}(\cdot)$  of the true curve $X_i(\cdot)$. We again use the PVE criterion to select the number of PC. So for sparse data observed on irregular grid and observed with measurement error, we employ FPCA to get $\hat{X_{i}}(t)$ and then use \{$Y_i(t_{ij}),\hat{X_{i}}(t_{ij}),  j=1,2,\ldots,m_i\}_{i=1}^{n}$ as our original data to perform our proposed test. Our simulations show that we are able to maintain correct type I error rate and obtain satisfactory power of our testing method using this strategy.
\vspace{-3 mm}
\subsection{Extension to multiple covariates}
\label{subsec:multcov}
Now we illustrate how our testing method can be applied in multiple covariate setting. 
\newpage
\hspace*{ -5 mm}
In this case the likelihood is as in (2.4) given by
\begin{equation}
L_{ML}(\tau_0,\tau_1,\ldots,\tau_p)=-1/2(\ln|\^V|+\*Y^{T}\^V^{-1}\*Y),
\end{equation}
with the only difference being $\^V=\^V(\tau_0,\tau_1,\ldots,\tau_p)=\#\Sigma+\tau_0\^B\^B^{T}+\tau_1\^Z_1\^Z_1^{T}+\ldots+\tau_p\^Z_p\^Z_p^{T}$. So testing for the effect of $X_p(\cdot)$ here similarly reduces to testing for the variance component $\tau_p$, namely, we test $H_{0} :\tau_p =0$ vs $H_{1} : \tau_p> 0 $. The score function $S_{\tau_p}(\tau_0,\tau_1,\ldots,\tau_p)$ and the information matrix $\^I(\bm\theta)$ \{$\bm\theta=(\tau_0,\tau_1,\ldots,\tau_p)^T$\} are also modified accordingly taking into account the additional variance components, e.g., the information matrix $\^I(\bm\theta)$ now has to be partitioned as  
\[
 \^I(\bm\theta)=  \begin{Bmatrix} 
      \^I_{11}(\bm\theta)_{p\times p} &  \*I_{12}(\bm\theta)_{p\times 1}  \\ 
      \*I_{21}(\bm\theta)^{T}_{1\times p} & I_{22}(\bm\theta)_{1\times 1}  
      \end{Bmatrix}.
\]
Then our testing method can applied as in Section \ref{subsec:Test_m} and the generalization of Theorem (2.1) and Theorem (2.2) to this multicovariate setting remains valid. We have given examples of application of our testing method in multicovariate setting, both in our simulation study in Section \ref{sec:sim_stud} and real data application in Section \ref{sec:real_dat}, where we have considered two covariate scenarios.
\section{Simulation Study}
\label{sec:sim_stud}

\subsection{Study design}
In this section, we investigate the performance of our testing method via simulation study. We evaluate our test in terms of type I error rate and power. We also compare our method to the existing bootstrapped F-test proposed by \cite{kim2016general}. To this end, we consider the following two scenarios.

\hspace*{- 6 mm}
\textbf{Scenario A (single covariate)}: 
We generate data from the model,
$$Y_i(t)=\beta_{0}(t)+X_{i}(t)\beta_{1}(t)+\epsilon_{i}(t),$$ 
where $\beta_{0}(t)= 1+2t+t^2$ and $\beta_{1}(t)= t/8$. The original covariate $X_{i}(\cdot)$ are i.i.d. copies of $X(\cdot)$, where 
$X(t)=a+b\hspace{2mm} \sqrt[]{2}sin(\pi t)+ c\hspace{2mm} \sqrt[]{2}cos(\pi t)$, where $a\sim \mathcal{N}(0,1)$, $b\sim \mathcal{N}(0,.85^2)$ and $c\sim \mathcal{N}(0,.70^2)$ and they are independent. As discussed in Section \ref{sec:method}, we assume that we observe $X_{i}(t)$ with measurement error, i.e., we observe $U_{i}(t)=X_{i}(t)+\delta$, where $\delta\sim \mathcal{N}(0,.6^2)$. The error process $\epsilon_{i}(t)$ is generated as 
$$\epsilon_i(t)=\xi_{i1}\hspace{2mm} \sqrt[]{2}cos(\pi t) +\xi_{i2} \hspace{2mm}\sqrt[]{2}sin(\pi t) + N(0,0.9^2I_{m_{i}}), $$
where $\xi_{i1}\stackrel{iid}{\sim}\mathcal{N}(0,2)$  and $\xi_{i2}\stackrel{iid}{\sim}\mathcal{N}(0,0.75^2)$. We consider the following sampling designs:
\begin{itemize}
\item Dense design:
Functional data are observed in $S$ for each subject, where $S$ is the set of $m = 81$ equidistant time points in $\mathcal{T}=[0,1].$ 
\item Sparse design: The response $Y_i(t)$ and noisy covariate
$U_{i}(t)$ both are observed in random $m_{Y_i}$ and $m_{U_i}$ points in $S$ where $m_{Y_i}\stackrel{iid}{\sim} Uniform\{20,21,\ldots,31\}$ and also $m_{U_i}\stackrel{iid}{\sim} Uniform\{20,21,\ldots,31\}.$ 
\end{itemize}
We consider two sample sizes, $n=100$ and $300$ for comparison with the bootstrapped F test method. An additional simulation is done in this same set up for dense data with $m=20$, $n=39$ to illustrate the performance of the testing method for finite sample sizes, as considered in the gait analysis in later section of this paper. \\
\textbf{Scenario B (Multiple Covariates)}:\\
We generate data from the model,
$$Y_i(t)=\beta_{0}(t)+X_{i1}(t)\beta_{1}(t)+X_{i2}(t)\beta_{2}(t)+\epsilon_{i}(t),$$
where $\beta_{0}(t)= 1+2t+t^2$, $\beta_{1}(t)= t/8$ and $\beta_{2}(t)=sin(\pi t) $. The original covariates
$X_{ik}(\cdot)$ are i.i.d. copies of $X_k(\cdot)$, where 
$X_k(t)=a_k+b_k\hspace{2mm}\sqrt[]{2}sin(\pi t)+ c_k\hspace{2mm}\sqrt[]{2}cos(\pi t)$, where $a_k\sim \mathcal{N}(0,(2^{-.5(k-1)})^2)$, $b_k\sim \mathcal{N}(0,(0.85\times 2^{-.5(k-1)})^2)$, and  $c_k\sim \mathcal{N}(0,(0.70\times 2^{-.5(k-1)})^2)$, and they are independent. The covariates $X_k(\cdot)$ are same as considered in \cite{kim2016general}. We observe $X_{ik}(t)$ with measurement error, i.e., we observe $U_{ik}(t)=X_{ik}(t)+\delta_k$, where $\delta_k\sim \mathcal{N}(0,.6^2)$. The error process $\epsilon_{i}(t)$ are generated as in Scenario A described above. Similar sparse and dense design settings and sample sizes $n\in\{100,300\}$ are considered. Here the main goal is to test $H_{0} : \beta_2(t)=0 $ against the alternative $H_{1} : \beta_2(t)\neq 0 $. For each of the scenarios we use 1000 generated data sets to asses type I error and power. To model the regression functions, 7 cubic B-splines with equally spaced knots are used for both simulation scenarios.
\subsection{Simulation Results}
\textbf{Scenario A}:\\
We first assess type I error of the test. We use nominal levels of $\alpha=5\%$ for $n=100$ and $n=300$. The results are displayed in Table \ref{my-label}. The estimated standard errors are also given.\\
\hspace*{55 mm} [Table 1 about here]\\
We observe that our test maintains nominal type I error rate. For dense design, the estimated type I error lies below the nominal level of $.05$. For sparse design, As sample size increases, the size performance of the test improves, which is expected as the proposed test is a large sample one and for the asymptotic convergence to the null distribution to hold we need larger sample size. The nominal level lies within two standard error limit of estimated type I error in this case.

Next, we study the power performance of our test for a fixed nominal level $\alpha= 5\%$. To this end, we generate data in the simulation set up mentioned earlier with $\beta_{1}(t)= dt/8$. Then $d=0$ corresponds to the null hypothesis and $d>0$ plays the role of effect size and captures the departure from null hypothesis. We compare the power of our test to the bootstrapped-F test for $n\in\{100,300\}$, and for both dense and sparse sampling designs. For comparison of power with the  bootstrapped method, we use the results from the simulation study conducted by \cite{kim2016general}. Results of our study are displayed in Figure \ref{fig:fig4}.\\
\hspace*{55 mm} [Figure 1 about here]\\
We note that across the sparse and dense design scenarios, the proposed score based test produces higher power than the bootstrapped-F test method. This is expected as our method is a likelihood based method and directly uses value of the score. We also observe that as the sample size increases or $d$ increases the power of our test converges to one across all the settings faster than the bootstrapped-F test. Similar results for the additional simulation study is provided in the Figure S1 of supplemental material.\\
\textbf{Scenario B}:\\
 The type I error rates for this scenario are also displayed in Table \ref{my-label}. Nominal levels of $\alpha=5\%$ for $n=100$ and $n=300$ are considered. Again we observe the proposed test maintaining nominal type I error rate, and improvement in size performance with increasing sample size, particularly for sparse design. A similar evaluation of power performance is done as in Scenario A. Namely we generate data as in simulation Scenario B with $\beta_{2}^*(t)= d\beta_2(t)$, where $d$ is a constant. The power curve of our testing method for both dense and sparse settings for $n\in\{100,300\}$ are displayed in Figure \ref{fig:figb}. We observe the power converging to one, as sample size or $d$ increases, across all the sampling designs.\\
 \hspace*{55 mm} [Figure 2 about here]\\
 
 Our simulation results illustrate the proposed testing method is able to maintain the nominal type I error rate as well as capture the departure from null hypothesis successfully, even when data is observed sparsely and there are multiple covariates observed with measurement error.
  
 We have used $7$ cubic B-splines basis functions to model the regression functions in our simulations. The choice of the number of basis in reality depends on the type of design (dense or sparse) and number of observed time points for each subjects. To illustrate the effect of number of basis on the performance of the test an additional simulation is carried out as in Scenario A for dense design and sample size $n=300$, with number of basis $k_1= 7,10,12,14,16$. The result is illustrated in Figure S2 of supplemental material. The estimated type I error of the test maintain the nominal level for all the cases. Noticeably, we observe a marginal decrease in the power with a larger number of basis functions, particularly for smaller effect size, which might be attributed to more random effects in the model and the coefficient functions getting more rough with the increasing number of basis. Analogous comparison for Scenario B is provided in Figure S3 of supplemental material where similar trend can be observed.
\section{Real Data Applications}
\label{sec:real_dat}
Through simulations we have shown our proposed testing method is able to identify significant covariates under both dense and sparse sampling design, even when the covariates are observed with measurement error. Next we present two real data applications of our testing method to demonstrate its usefulness in identifying significant time varying covariates in practical problems. We first consider the study of gait deficiency which is a typical case of dense data with small measurement error, subsequently we also apply our method to a study of dietary calcium absorption, where the data is sparse and measurement error is relatively higher.  
\subsection{Gait Data}
 In this study, the goal is to understand how the joints in hip and knee interact during a gait cycle \citep{theologis2009children}, \citep{Ramsay05functionaldata}. Here, there are longitudinal measurements of hip and knee angles taken on 39 children on 20 equispaced evaluation points in $\mathcal{T}=[0,1]$. Figure \ref{fig:webfig4} displays the observed individual trajectories of the hip and knee angles.\\
 \hspace*{55 mm} [Figure 3 about here]\\
 The study of gait is important as it helps to identify issues causing pain, and also implement and evaluate treatments to correct abnormalities.
 As discussed earlier, one natural question to ask here is whether the knee angles (response) are at all associated with the hip angles (covariate). In our terminology, here $Y_i(t)$ is the knee angle. We assume the hip angles $X_{i}(t)$ are observed with measurement error, i.e, we observe $U_{i}(t_{ij})=X_{i}(t_{ij})+\delta_{ij}$,
where $\delta_{ij}$ are assumed to be white noise. 
We use the functional linear concurrent model (1) in this paper to model the time varying effect of hip angles on knee angles. The FLCM can be see as an extension of classical regression model allowing the covariates to have dynamic effects. In the functional linear concurrent modeling setup, we are therefore interested in testing $H_{0} : \beta_1(t)=0 $ against the alternative $H_{1} : \beta_1(t)\neq 0 $, where $\beta_1(t)$ denotes the linear concurrent effect of hip angles on knee angles. We use the proposed score test method with 7 cubic B-splines to model the time varying effect $\beta_{1}(t)$. The p-value of the proposed test is calculated to be $<10^{-4}$. So we reject the null hypothesis and conclude that knee angle at any fixed time point is associated with the hip angle at the same time point. Our findings match with that of \cite{kim2016general}, who found the association to be significant using a bootstrapped-F test.

The effect of the number of basis function on the test is investigated by varying the number of cubic B-splines basis functions between $5$ to $10$, in all cases the p-value of the proposed test was calculated to be $<10^{-2}$, further confirming existence of significant concurrent association between knee and hip angles.

As the sample size ($n=39$) for this data is small and our testing method is an asymptotic one, we further evaluate the performance of our proposed method using a simulation study that captures the feature of the gait data. This also enables us to see the power performance of our method. This is done in a similar way as that of \cite{kim2016general}. We use a model that mimic the feature of the gait data, generate a large simulated data set and assess the power performance of our method on the simulated data. In particular, We generate the covariate $X_{i}(t)$ from a process with the mean and covariance functions that equal their estimated counterparts from the data using FPCA. We also estimate the parameters $\beta_{0}(t)$, $\beta_1(t)$ from the fit of our full model and estimate $\Sigma(s,t)$ by doing FPCA on the residuals obtained from the full model fit. Subsequently we generate observations using the model $Y_i(t)=\hat{\beta_{0}}(t)+d\{\hat{X_{i}}(t)\hat{\beta_{1}}(t)\}+\epsilon_{i}(t)$, where $d$ is a constant, and $\epsilon_{i}(\cdot){\sim} \mathcal{N}(0,\hat{\Sigma}(\cdot,\cdot))$. We simulate 
$n=300$ response curves $Y_i(t)$ from the above set up. In the above set up $d$ again plays the role of a parameter, which controls departure from null hypothesis. We perform a power analysis simulating 1000 such data sets from above scenario for various $d$. Figure \ref{fig:fig7} shows the power curve obtained from our analysis using the proposed testing method with 7 cubic B-spline basis functions.\\
\hspace*{55 mm} [Figure 4 about here]\\
For $d=0$, the type I error is $.062$, and nominal level $\alpha=.05$ is within its 2-standard errors limit. As $d$ increases the power gradually increases and ultimately goes to one
when $d=0.1$, which ensures power is one at $d=1$. Similar power curves with different number of basis functions are also provided in Figure \ref{fig:fig7}. For a small number basis functions (5) we observe slightly inflated type I error, for larger number of basis (9), a loss in power is noticed for small effect sizes. Again the best results are obtained using a moderate number of basis functions. The results further confirm our conclusion that knee angle made during a gait cycle is concurrently associated with hip angle.
\vspace*{- 5mm}
\subsection{Calcium Absorption Data}
We consider the dietary calcium absorption study given in \cite{davis2002statistical}. In this study the subjects are a group of 188 patients. We have data on calcium absorption, dietary calcium intake and BMI of these patients at irregular intervals between 35 to 64 years of their ages. The number of repeated measurements for each patient is between 1 to 4. Figure S4 in supplemental material shows the individual curves of patients' calcium absorption and calcium intake along their ages. We notice that this is a typical case of sparse data with relatively high measurement error. In this study, we are interested in finding whether calcium intake has any effect on calcium absorption in presence of BMI. We assume the covariates BMI $X_{i1}(t)$ and calcium intake $X_{i2}(t)$ are observed with measurement error. So our observed data is $U_{i}(t_{ij})=X_{i1}(t_{ij})+\delta_{ij}$ and $V_{i}(t_{ij})=X_{i2}(t_{ij})+\nu_{ij}$. Our response variable is  calcium absorption $Y_i(t)$.  Following the study in \cite{kim2016general} we assume a functional linear concurrent regression model, 
$Y_i(t)=\beta_{0}(t)+X_{i1}(t)\beta_{1}(t)+ X_{i2}(t)\beta_{2}(t)+\epsilon_{i}(t)$, and want to test $H_{0} : \beta_{2}(t)=0 $ against the alternative $H_{1} : \beta_{2}(t)\neq 0.$

First we use FPCA as discussed in Section $\ref{subsec:ext}$ on noisy covariates to get their smooth counterparts at time points where we have $Y_i(t)$ available and then apply our one sided score test method for multiple covariates described as in Section \ref{subsec:multcov}. We use 7 cubic B-Splines to model $\beta_{0}(t)$, $\beta_{1}(t)$ and $\beta_{2}(t)$. The p-value of our test is calculated to be $<10^{-5}$. Thus we conclude in presence of BMI, calcium intake of patients has a significant effect on calcium absorption which again matches with the findings in \cite{kim2016general} and other studies of dietary calcium absorption.

\section{Discussion and Future Work}
\label{sec disc}
In this article, we have proposed a likelihood based method for testing of hypothesis in functional linear concurrent regression. We have formulated the problem as a test for variance component and have used a one sided score test approach. We have established the asymptotic null distribution of our test statistic under some standard assumptions. Through simulations we have shown our proposed method maintains the nominal type I error rate and also yield higher power compared to the existing bootstrapped-F test, even when data is observed sparsely and with measurement error. 
We have successfully applied our method in finding significance of covariates in two real data application, namely the gait study and calcium absorption study. We note our method is a general one and can be applied for testing in longitudinal data setting too, where we have more flexibility in assuming parametric form of error covariance structure and estimating it consistently from the data.
%

We have considered approximating the true coefficient function using finite cubic B-spline basis expansion. We have used cubic B-spline basis with equispaced knots for computational tractability. The placement and choice of optimal number of knots is itself a challenge and open problem in functional data and there is no single consensus \citep{Ramsay05functionaldata,vsevolozhskaya2014functional}. Even if the true function lies in infinite dimensional space we have shown the proposed test performs well with a moderate number of basis functions. The choice of number of basis functions is subjective and depends on the type of design and number of observed time points for each subject. We have demonstrated using a large number of basis functions leads to loss of power of the test and therefore using a moderate number of basis functions is recommended. 

We have considered Gaussian distribution for the functional error in our model, if the distribution is non Gaussian, one can still use the score test statistic proposed in this paper and employ a subject level bootstrap for performing the test. Similarly it would be of interest to explore how the score test approach can be extended to generalized functional concurrent model such as logistic regression. In developing our test, we have used a random effects formulation of the problem arising from directly penalizing the coefficient functions. It is also plausible to use penalty on $r$-th ($r>0$) derivative of the coefficient functions. In this case the main challenge is to handle the singularity of the resulting covariance matrices, which can be addressed by using mixed effects model and subsequently testing for variance component using variants of likelihood ratio tests \citep{crainiceanu2004likelihood}. Another important work for future could be to prove the consistency results of the FPCA approximation methods used in this article. Further we would like to extend our testing method to nonparametric functional concurrent regression and more general function on function regression models and these are possible areas for future research that could explored based on our work.
\section*{Software}
\label{sec9}
Software in the form of R code, together with the data set and complete documentation is available at GitHub (\url{https://github.com/rahulfrodo/FLCM_Score}).

\section*{Acknowledgements}
We would like to thank Dr. Janet Kim and Dr. Ana-Maria Staicu for sharing the results of their method \citep{kim2016general} with us.

\section*{Supplemental Material}
Appendix S1, Appendix S2 referenced in Section~\ref{subsec:Test_m}, and Figures S1$-$S4 are available online with Supplemental Material.



\bibliographystyle{asa}
\bibliography{interactapasample}

\begin{thebibliography}{29}
\newcommand{\enquote}[1]{``#1''}
\expandafter\ifx\csname natexlab\endcsname\relax\def\natexlab#1{#1}\fi

\bibitem[{Cai et~al.(2000)Cai, Fan, and Yao}]{cai2000functional}
Cai, Z., Fan, J., and Yao, Q. (2000), \enquote{Functional-coefficient
  regression models for nonlinear time series,} \textit{Journal of the American
  Statistical Association}, 95, 941--956.

\bibitem[{Crainiceanu and Ruppert(2004)}]{crainiceanu2004likelihood}
Crainiceanu, C.~M. and Ruppert, D. (2004), \enquote{Likelihood ratio tests in
  linear mixed models with one variance component,} \textit{Journal of the
  Royal Statistical Society: Series B (Statistical Methodology)}, 66, 165--185.

\bibitem[{Davis(2002)}]{davis2002statistical}
Davis, C.~S. (2002), \enquote{Statistical methods for the analysis of repeated
  measurements,} .

\bibitem[{Fan and Zhang(2000)}]{fan2000simultaneous}
Fan, J. and Zhang, W. (2000), \enquote{Simultaneous confidence bands and
  hypothesis testing in varying-coefficient models,} \textit{Scandinavian
  Journal of Statistics}, 27, 715--731.

\bibitem[{Gelfand et~al.(2003)Gelfand, Kim, Sirmans, and
  Banerjee}]{gelfand2003spatial}
Gelfand, A.~E., Kim, H.-J., Sirmans, C., and Banerjee, S. (2003),
  \enquote{Spatial modeling with spatially varying coefficient processes,}
  \textit{Journal of the American Statistical Association}, 98, 387--396.

\bibitem[{Greven et~al.(2008)Greven, Crainiceanu, K{\"u}chenhoff, and
  Peters}]{greven2008restricted}
Greven, S., Crainiceanu, C.~M., K{\"u}chenhoff, H., and Peters, A. (2008),
  \enquote{Restricted likelihood ratio testing for zero variance components in
  linear mixed models,} \textit{Journal of Computational and Graphical
  Statistics}, 17, 870--891.

\bibitem[{Guo(2002)}]{guo2002functional}
Guo, W. (2002), \enquote{Functional mixed effects models,} \textit{Biometrics},
  58, 121--128.

\bibitem[{Hall et~al.(2006)Hall, M{\"u}ller, Wang, et~al.}]{hall2006properties}
Hall, P., M{\"u}ller, H.-G., Wang, J.-L., et~al. (2006), \enquote{Properties of
  principal component methods for functional and longitudinal data analysis,}
  \textit{The annals of statistics}, 34, 1493--1517.

\bibitem[{Hastie and Tibshirani(1993)}]{hastie1993varying}
Hastie, T. and Tibshirani, R. (1993), \enquote{Varying-coefficient models,}
  \textit{Journal of the Royal Statistical Society. Series B (Methodological)},
  757--796.

\bibitem[{Huang et~al.(2002)Huang, Wu, and Zhou}]{huang2002varying}
Huang, J.~Z., Wu, C.~O., and Zhou, L. (2002), \enquote{Varying-coefficient
  models and basis function approximations for the analysis of repeated
  measurements,} \textit{Biometrika}, 89, 111--128.

\bibitem[{Huang et~al.(2004)Huang, Wu, and Zhou}]{huang2004polynomial}
--- (2004), \enquote{Polynomial spline estimation and inference for varying
  coefficient models with longitudinal data,} \textit{Statistica Sinica},
  763--788.

\bibitem[{Kim et~al.(2018)Kim, Maity, and Staicu}]{kim2016general}
Kim, J., Maity, A., and Staicu, A.-M. (2018), \enquote{Additive nonlinear
  functional concurrent model,} \textit{Statistics and Its Interface}, 11,
  669--685.

\bibitem[{Li et~al.(2010)Li, Hsing, et~al.}]{li2010uniform}
Li, Y., Hsing, T., et~al. (2010), \enquote{Uniform convergence rates for
  nonparametric regression and principal component analysis in
  functional/longitudinal data,} \textit{The Annals of Statistics}, 38,
  3321--3351.

\bibitem[{Liang and Self(1996)}]{liang1996asymptotic}
Liang, K.-Y. and Self, S.~G. (1996), \enquote{On the asymptotic behaviour of
  the pseudolikelihood ratio test statistic,} \textit{Journal of the Royal
  Statistical Society: Series B (Methodological)}, 58, 785--796.

\bibitem[{Lin(1997)}]{lin1997variance}
Lin, X. (1997), \enquote{Variance component testing in generalised linear
  models with random effects,} \textit{Biometrika}, 84, 309--326.

\bibitem[{Mercer(1909)}]{j1909xvi}
Mercer, J. (1909), \enquote{Functions of positive and negative type, and their
  connection with the theory of integral equations,} \textit{Phil. Trans. R.
  Soc. Lond. A}, 209, 415--446.

\bibitem[{Molenberghs and Verbeke(2007)}]{molenberghs2007likelihood}
Molenberghs, G. and Verbeke, G. (2007), \enquote{Likelihood ratio, score, and
  Wald tests in a constrained parameter space,} \textit{The American
  Statistician}, 61, 22--27.

\bibitem[{Ramsay and Silverman(2005)}]{Ramsay05functionaldata}
Ramsay, J. and Silverman, B. (2005), \enquote{Functional Data Analysis,} .

\bibitem[{Self and Liang(1987)}]{self1987asymptotic}
Self, S.~G. and Liang, K.-Y. (1987), \enquote{Asymptotic properties of maximum
  likelihood estimators and likelihood ratio tests under nonstandard
  conditions,} \textit{Journal of the American Statistical Association}, 82,
  605--610.

\bibitem[{{\c{S}}ent{\"u}rk and Nguyen(2011)}]{csenturk2011varying}
{\c{S}}ent{\"u}rk, D. and Nguyen, D.~V. (2011), \enquote{Varying coefficient
  models for sparse noise-contaminated longitudinal data,} \textit{Statistica
  Sinica}, 21, 1831.

\bibitem[{Staicu et~al.(2014)Staicu, Li, Crainiceanu, and
  Ruppert}]{staicu2014likelihood}
Staicu, A.-M., Li, Y., Crainiceanu, C.~M., and Ruppert, D. (2014),
  \enquote{Likelihood ratio tests for dependent data with applications to
  longitudinal and functional data analysis,} \textit{Scandinavian Journal of
  Statistics}, 41, 932--949.

\bibitem[{Theologis(2009)}]{theologis2009children}
Theologis, T. (2009), \enquote{Children’s Orthopaedics and Fractures (Chapter
  6),} .

\bibitem[{Vsevolozhskaya et~al.(2014)Vsevolozhskaya, Zaykin, Greenwood, Wei,
  and Lu}]{vsevolozhskaya2014functional}
Vsevolozhskaya, O.~A., Zaykin, D.~V., Greenwood, M.~C., Wei, C., and Lu, Q.
  (2014), \enquote{Functional analysis of variance for association studies,}
  \textit{PLoS One}, 9, e105074.

\bibitem[{Wang et~al.(2018)Wang, Zhong, Cui, and Li}]{wang2018unified}
Wang, H., Zhong, P.-S., Cui, Y., and Li, Y. (2018), \enquote{Unified empirical
  likelihood ratio tests for functional concurrent linear models and the phase
  transition from sparse to dense functional data,} \textit{Journal of the
  Royal Statistical Society: Series B (Statistical Methodology)}, 80, 343--364.

\bibitem[{Wu et~al.(1998)Wu, Chiang, and Hoover}]{wu1998asymptotic}
Wu, C.~O., Chiang, C.-T., and Hoover, D.~R. (1998), \enquote{Asymptotic
  confidence regions for kernel smoothing of a varying-coefficient model with
  longitudinal data,} \textit{Journal of the American statistical Association},
  93, 1388--1402.

\bibitem[{Yao et~al.(2005)Yao, M{\"u}ller, and Wang}]{yao2005functional}
Yao, F., M{\"u}ller, H.-G., and Wang, J.-L. (2005), \enquote{Functional data
  analysis for sparse longitudinal data,} \textit{Journal of the American
  Statistical Association}, 100, 577--590.

\bibitem[{Zhang and Lin(2003)}]{zhang2003hypothesis}
Zhang, D. and Lin, X. (2003), \enquote{Hypothesis testing in semiparametric
  additive mixed models,} \textit{Biostatistics}, 4, 57--74.

\bibitem[{Zhang and Lin(2008)}]{zhang2008variance}
--- (2008), \textit{Variance Component Testing in Generalized Linear Mixed
  Models for Longitudinal/Clustered Data and other Related Topics}, pp. 19--36.

\bibitem[{Zhang and Chen(2007)}]{zhang2007statistical}
Zhang, J.-T. and Chen, J. (2007), \enquote{Statistical inferences for
  functional data,} \textit{The Annals of Statistics}, 35, 1052--1079.

\end{thebibliography}

\newpage
\appendix
\section{Proof of Theorem 2.1: }
\label{app}
Suppose that the conditions (a) and (b) of Theorem 1 hold, i.e., the null hypothesis $H_{0} :\tau_1 =0$ holds, $\bm\theta_0 =(\tau_0^{*},0)$ is the true value of $\bm\theta$ and $\#\Sigma$ is the true covariance matrix of residual vector $\mathbf{\mathcal{E}}$. Denote $\^V=\^V(\bm\theta_0,\#\Sigma)$, $\*Y\sim\mathcal{N}(0,\^V)$ under null. Then $\^W=\^V^{-1/2}\*Y\sim\mathcal{N}(0,\^I)$.
Now $S_{\tau_1}(\bm\theta_0)=-1/2\{tr( \^Z^{T}\^V^{-1}\^Z)-(\^V^{-1/2}\*Y)^{T}\^V^{-1/2}\^Z\^Z^{T}\^V^{-1/2}(\^V^{-1/2}\*Y)\}$.
Hence 
\begin{align*}
\frac{S_{\tau_1}^2(\bm\theta_0)}{\Lambda(\bm\theta_0)} &=\frac{\left[1/2\{tr( \^Z^{T}\^V^{-1}\^Z)-(\^V^{-1/2}\*Y)^{T}\^V^{-1/2}\^Z\^Z^{T}\^V^{-1/2}(\^V^{-1/2}\*Y)\}\right]^{2}}{\Lambda(\bm\theta_0)}\\
&=\frac{\left[1/2\{tr(\frac{\^Z^{T}\^V^{-1}\^Z}{n})-(\^V^{-1/2}\*Y)^{T}\frac{\^V^{-1/2}\^Z\^Z^{T}\^V^{-1/2}}{n}(\^V^{-1/2}\*Y)\}\right]^{2}}{\frac{\Lambda(\bm\theta_0)}{n^2}}\\
&=\frac{\left[1/2\{\sum_{\ell=1}^{k_{1}}\lambda_\ell-(\^W)^{T}(\frac{\^V^{-1/2}\^Z \^Z^{T}\^V^{-1/2}}{n})(\^W)\}\right]^2}{\Lambda_n(\bm\theta_0)}\hspace{1 mm}\\
&=\frac{\left[1/2\{\sum_{\ell=1}^{k_{1}}\lambda_\ell-(\^W)^{T}\^U\^D\^U^T(\^W)\}\right]^2}{\Lambda_n(\bm\theta_0)}.
\end{align*}
Now we use the spectral decomposition $\^V^{-1/2}\^Z \^Z^{T}\^V^{-1/2}/n=\^U\^D\^U^{T}$. Then we have 
\begin{align*}
\frac{S_{\tau_1}^2(\bm\theta_0)}{\Lambda(\bm\theta_0)} &=\frac{\left[1/2\{\sum_{\ell=1}^{k_{1}}\lambda_\ell-(\^U^{T}\^W)^{T}\^D(\^U^{T}\^W)\}\right]^2}{\Lambda_n(\bm\theta_0)}\\
&=\frac{\{1/2(\sum_{\ell=1}^{k_{1}}\lambda_\ell-\^X^{T}\^D\^X)\}^2}{\Lambda_n(\bm\theta_0)} \\
&=\frac{\{1/2(\sum_{\ell=1}^{k_{1}}\lambda_\ell-\sum_{\ell=1}^{k_{1}}\lambda_\ell x_{\ell}^{2})\}^2}{\Lambda_n(\bm\theta_0)}.
\end{align*}
This follows from the fact  $\^X=\^U^{T}\^W\sim\mathcal{N}(0,\^I)$ as $\^U$ is an orthogonal matrix and nonzero eigenvalues of $\^V^{-1/2}\^Z\^Z^{T}\^V^{-1/2}/n$ and $\^Z^{T}\^V^{-1}\^Z/n$ are same. Therefore we have shown $\frac{S_{\tau_1}^2(\bm\theta_0)}{\Lambda(\bm\theta_0)} \overset{d}{=}\frac{\{1/2(\sum_{\ell=1}^{k_{1}}\lambda_\ell  x_{\ell}^{2}-\sum_{\ell=1}^{k_{1}}\lambda_\ell)\}^2}{\Lambda_n(\bm\theta_0)}$ under null, which completes the proof of our Theorem 1, namely: \vspace*{- 6 mm} $$T_S(\bm\theta_0,\#\Sigma) = \frac{S_{\tau_1}^2(\bm\theta_0)}{\Lambda(\bm\theta_0)}I(S_{\tau_1}(\bm\theta_0)\geq 0)\overset{d}{=} (1/2)^2\frac{\left(\sum_{\ell=1}^{k_{1}}\lambda_{\ell} x_{\ell}^{2}-\sum_{\ell=1}^{k_{1}}\lambda_\ell\right)^2}{\Lambda_n(\bm\theta_0)}I(\sum_{\ell=1}^{k_{1}}\lambda_{\ell}x_{\ell}^{2}\geq \sum_{\ell=1}^{k_{1}}\lambda_\ell),$$
where $x_{\ell}\stackrel{iid}{\sim}\mathcal{N}(0,1)$ and $\lambda_\ell$ are eigenvalues of $\^Z^{T}\^V(\bm\theta_0,\#\Sigma)^{-1}\^Z/n$.
\clearpage
\label{lastpage}
\newpage
\renewcommand{\thetable}{\arabic{table}}

\setcounter{table}{0}
\begin{table}[ht]
\centering
\caption{Estimated type 1 error along with their standard error at $\alpha=5\%$ level.}
\label{my-label}
\begin{tabular}{|l|l|l|l|l|}
\hline
Simulation Scenario & \multicolumn{2}{c}{Scenario A} & \multicolumn{2}{c|}{Scenario B}  \\\hline
Sampling  Design & $n=100$ & $n=300$ & $n=100$ & $n=300$  \\[.1cm] \hline
Dense & .038 (.006) & .041 (.006) & .049 (.007) & .046 (.007) \\
Sparse & .064 (.008) & .047 (.007) & .059 (.007) & .042 (.006) \\[.3cm] \hline
\end{tabular}
\end{table}
\newpage
\renewcommand{\thefigure}{\arabic{figure}}

\setcounter{figure}{0}
\begin{figure}[ht]
\centering
\scalebox{0.8}{\includegraphics{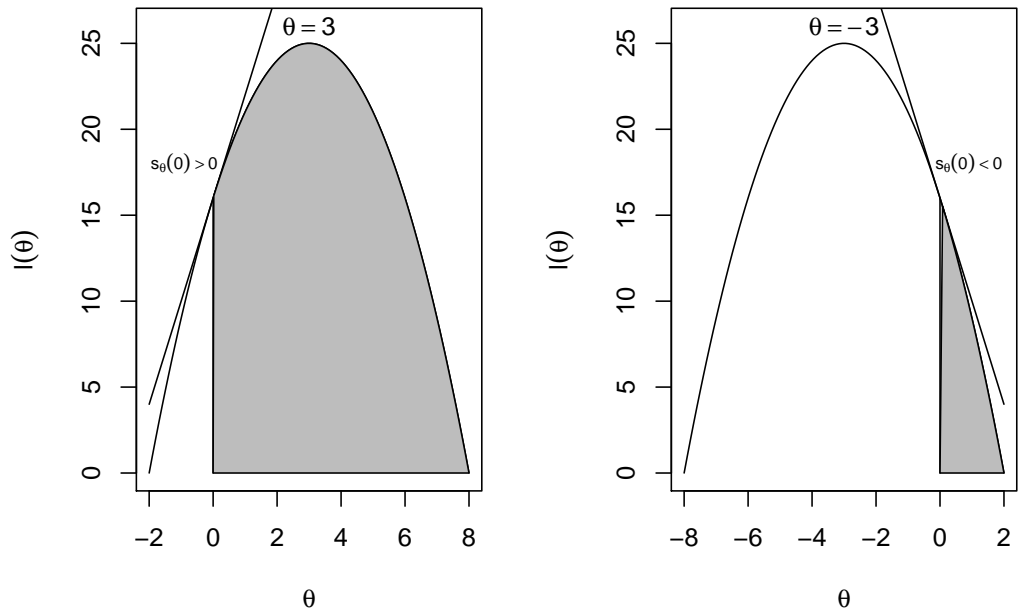}}\\
\caption{One sided score test for univariate case. Shaded area represents the alternative region. In the left panel the value of score at zero is positive giving evidence against null. In the right panel the value of score is negative and gives no evidence against null.}
\label{fig:fig0}
\end{figure}

\begin{figure}[ht]

\begin{center}
    \begin{tabular}{ll}
        \scalebox{0.4}{\includegraphics{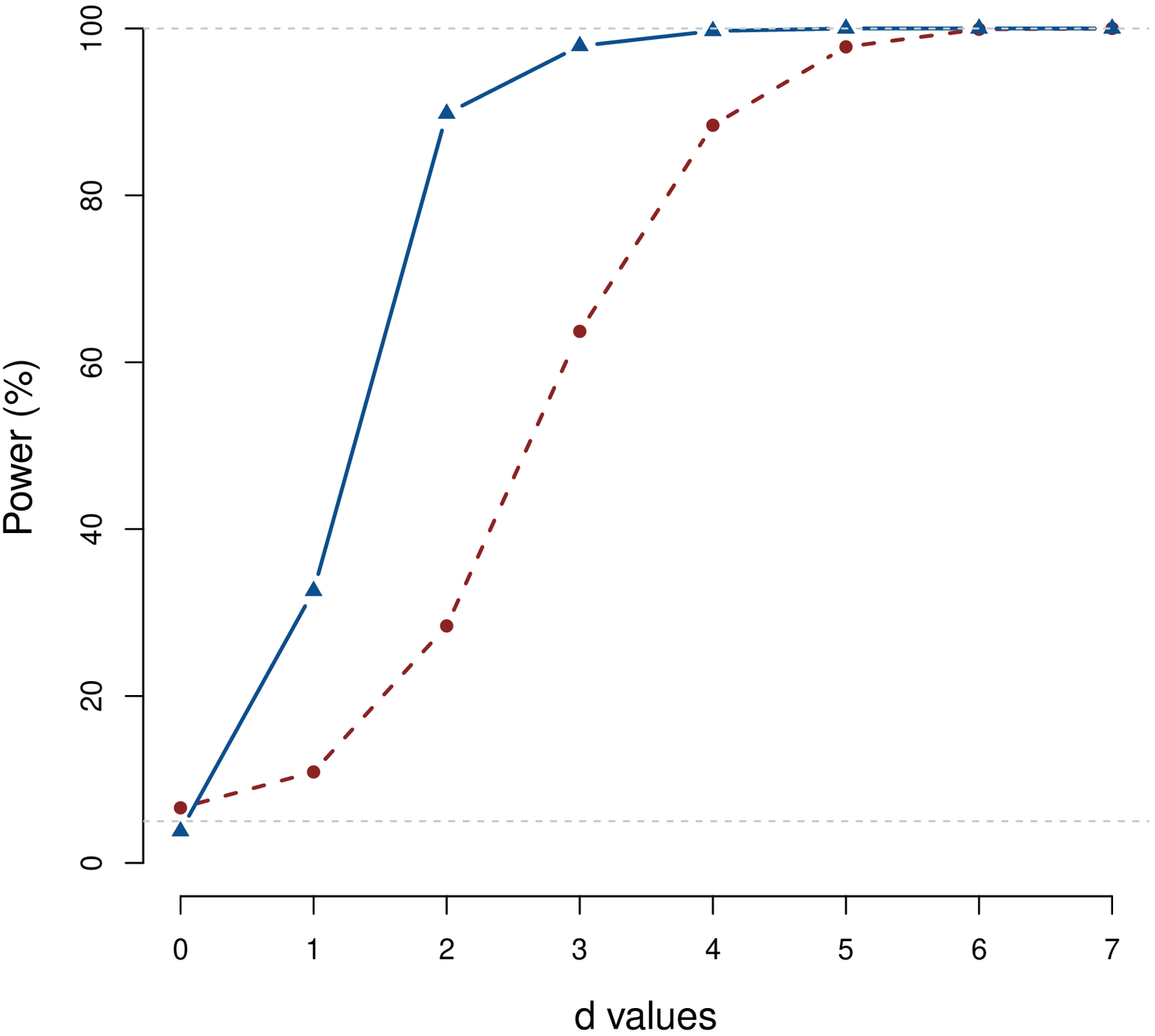}} &
 \scalebox{0.4}{\includegraphics{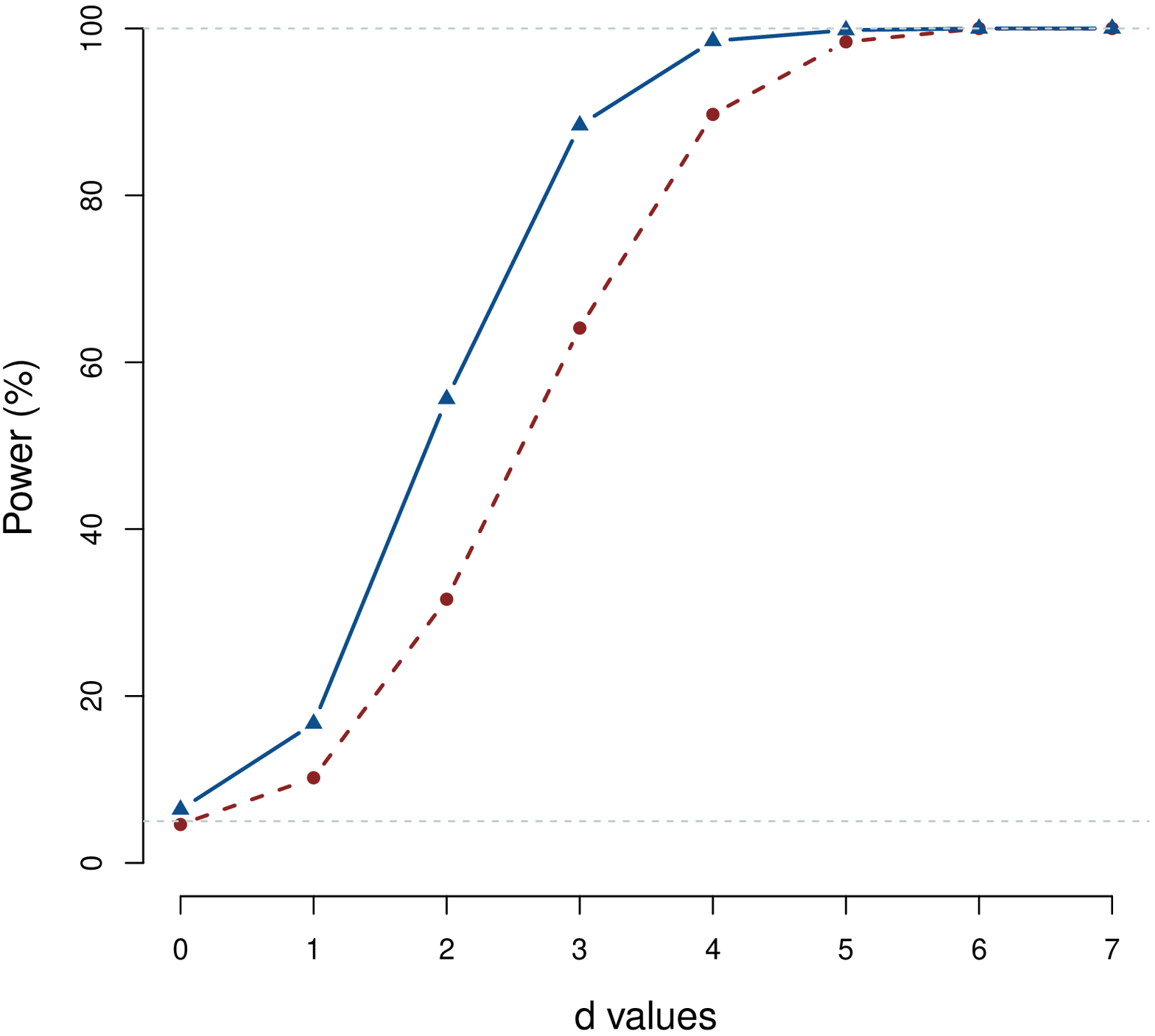}}\\
         \scalebox{0.4}{\includegraphics{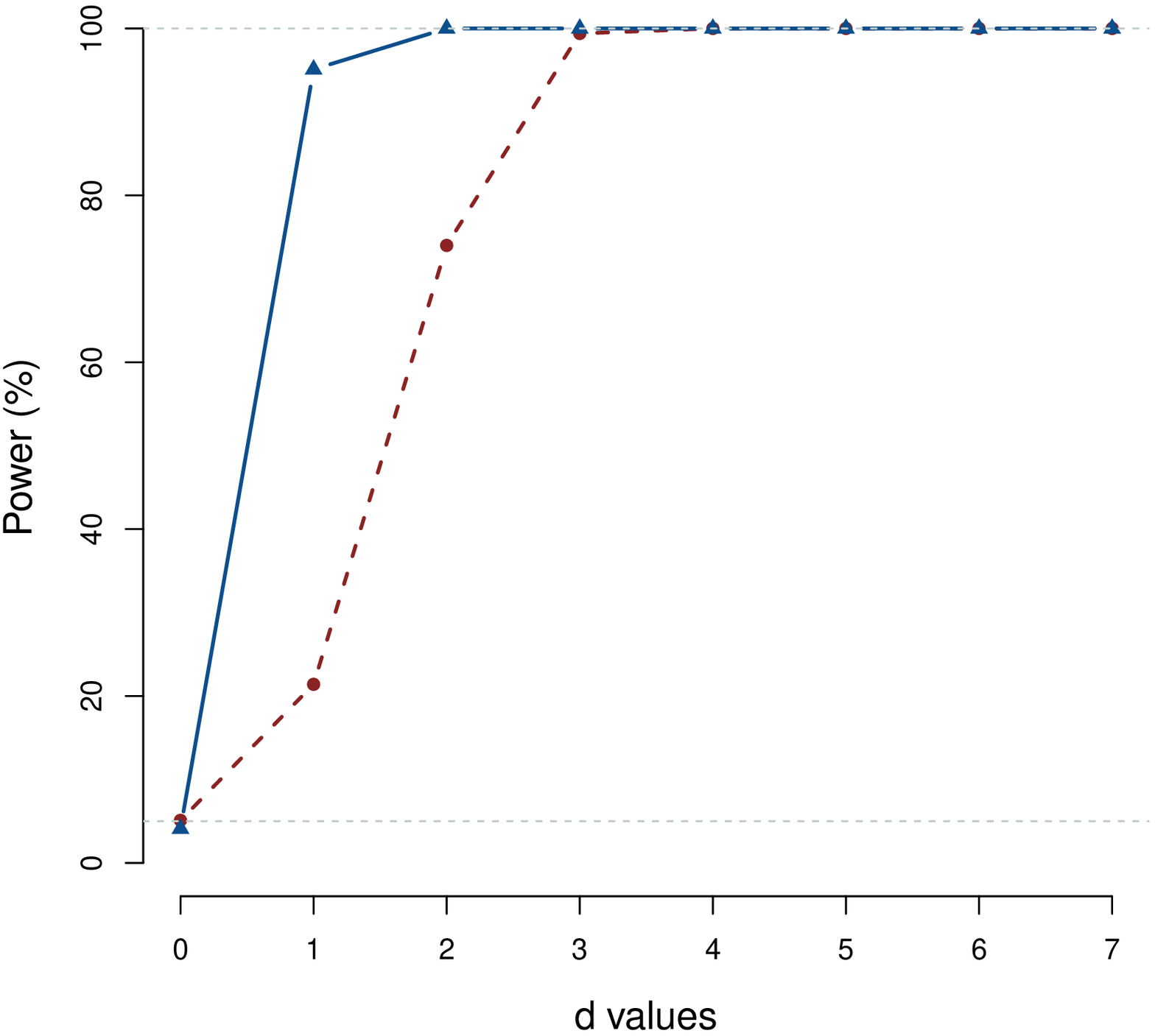}} &
 \scalebox{0.4}{\includegraphics{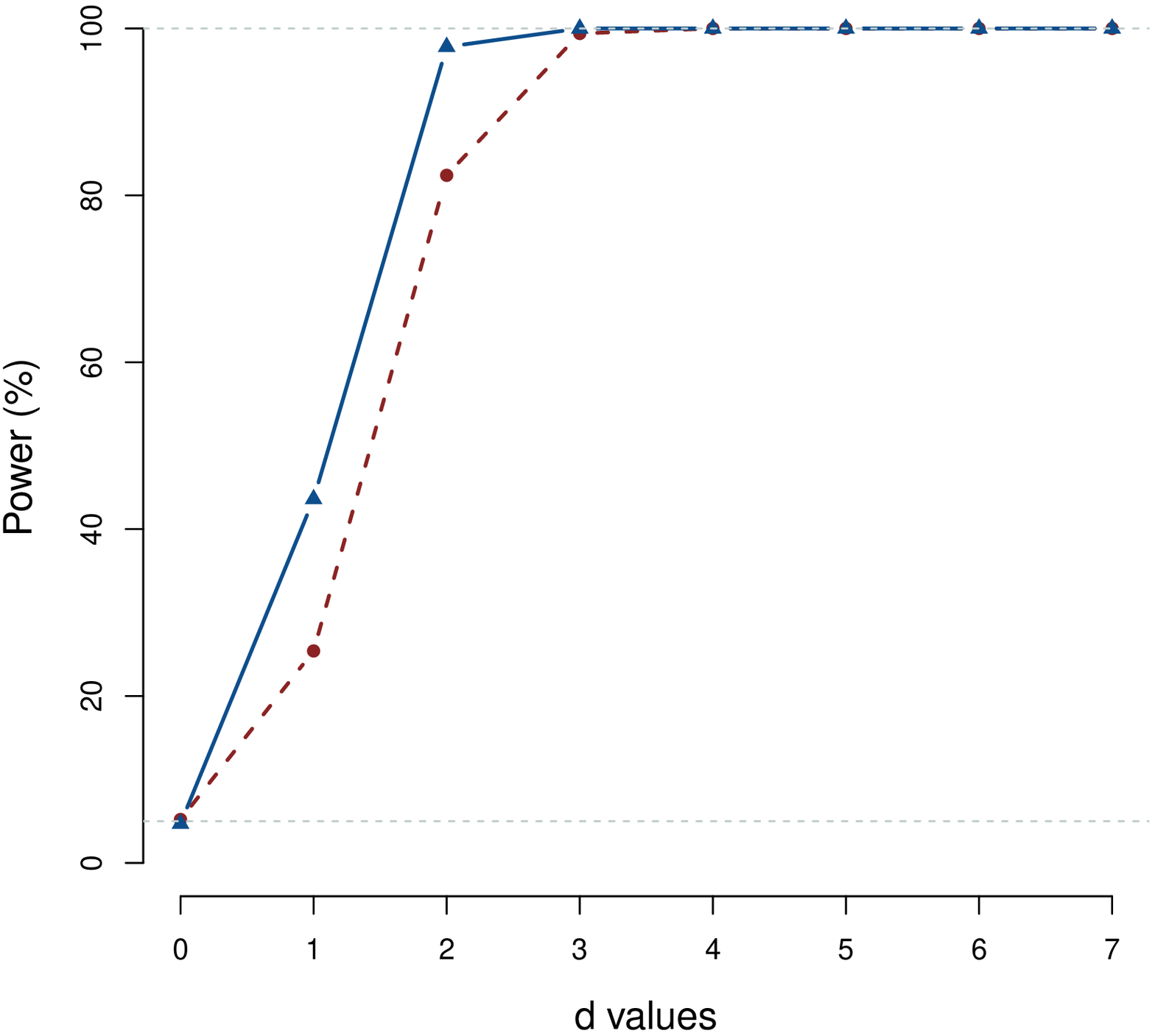}}\\
    \end{tabular}
\end{center}
\caption{Results of simulation study as described in Section \ref{sec:sim_stud}, Scenario A. Displayed are the power curves for our proposed procedure (solid line) and the bootstrap-F test based method (dashed line) for dense (left column) and sparse (right column) sampling designs with sample sizes $n = 100$ (top row) and $n = 300$ (bottom row).}
\label{fig:fig4}
\end{figure}

\begin{figure}[ht]
\begin{center}
\begin{tabular}{ll}
 \scalebox{0.4}{\includegraphics{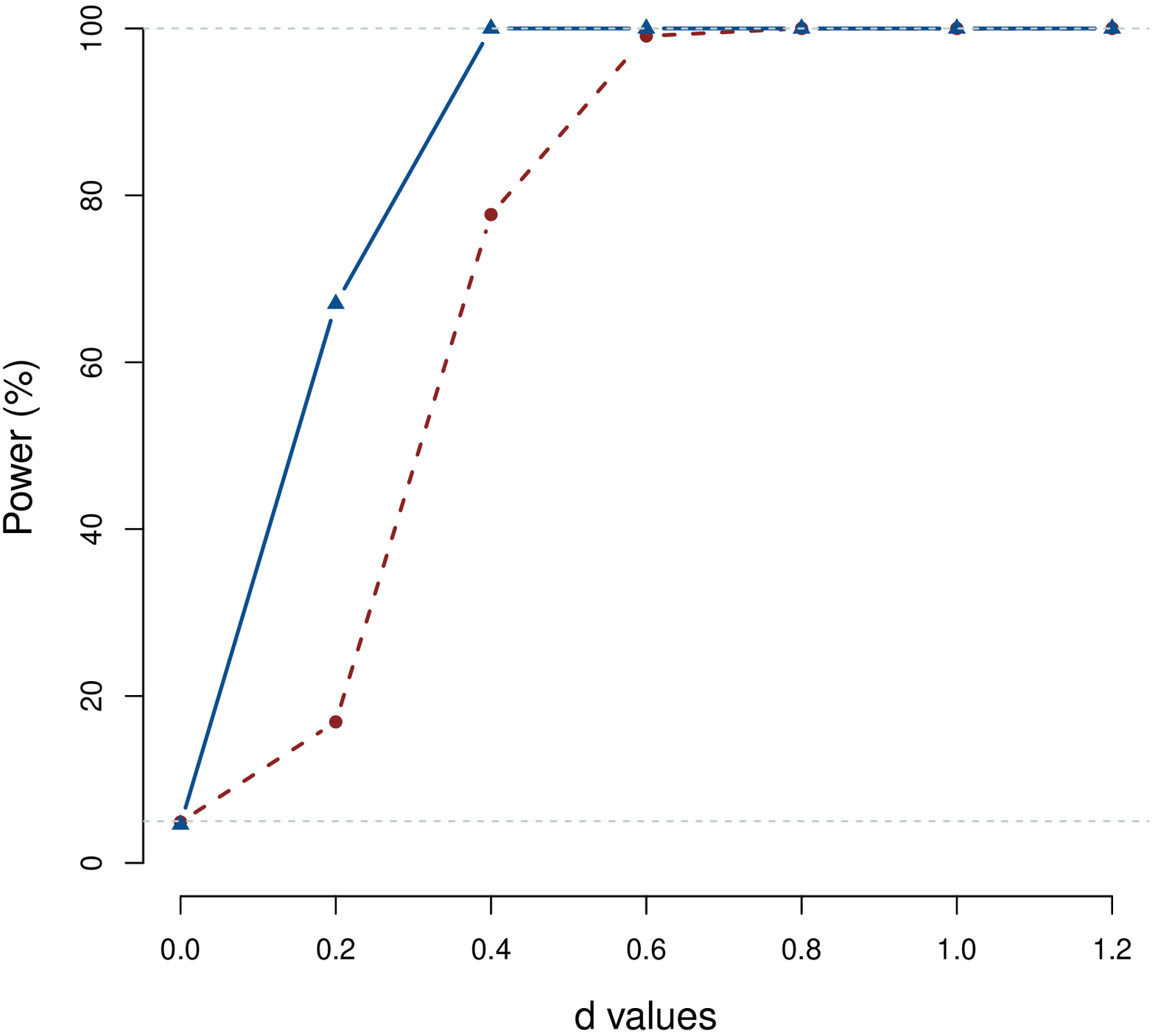}} &
 \scalebox{0.4}{\includegraphics{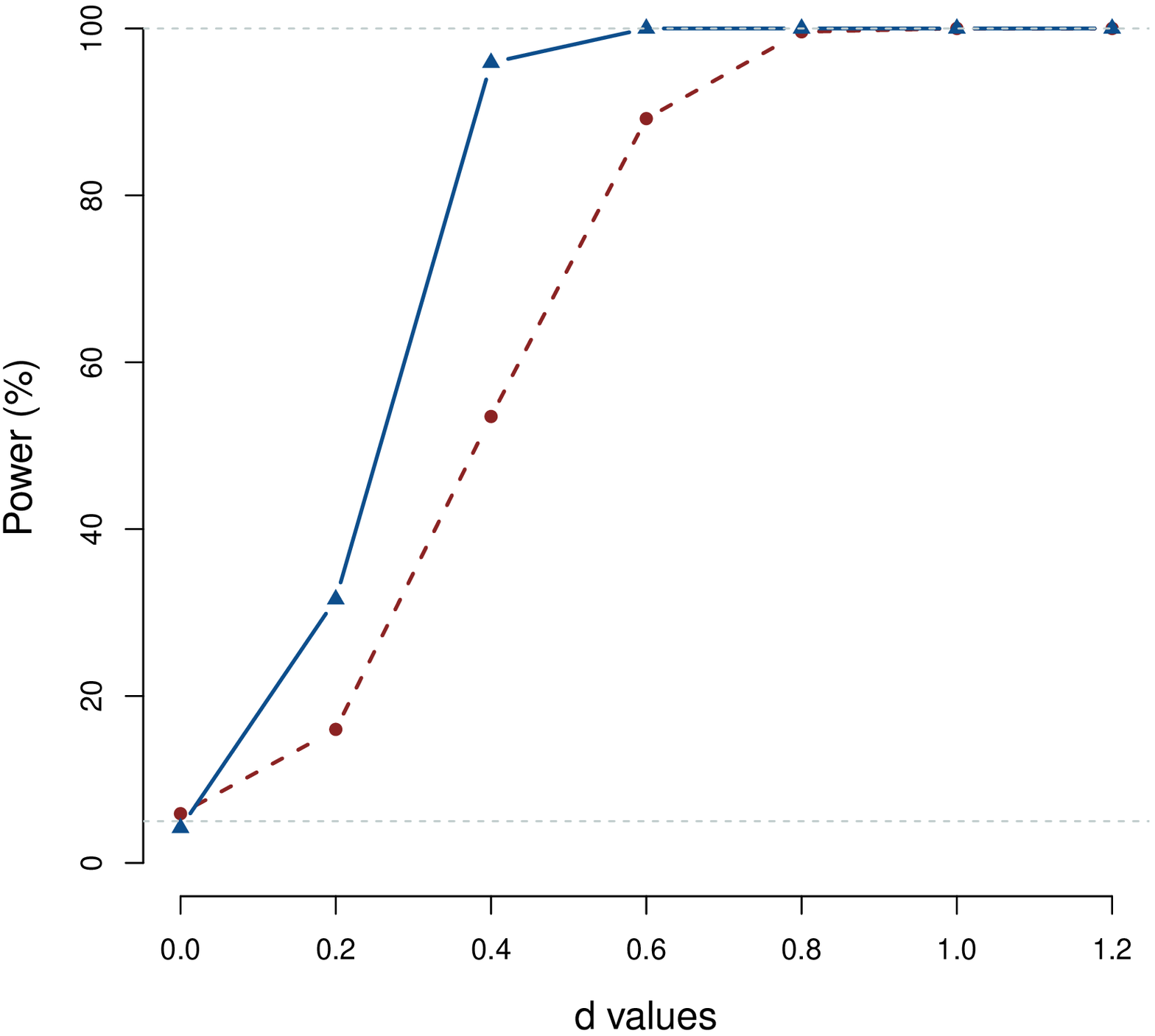}}\\
\end{tabular}
\end{center}
\caption{Results of simulation study as described in Section \ref{sec:sim_stud}, Scenario B. Displayed are the power curves for our proposed procedure with sample sizes $n = 100$ (dashed line) and $n = 300$ (solid line) for dense (left column) and sparse (right column) sampling designs.}
\label{fig:figb}
\end{figure}
\clearpage
\newpage

\begin{figure}[ht]
\centering
\includegraphics[width=.9\linewidth , height=.65\linewidth]{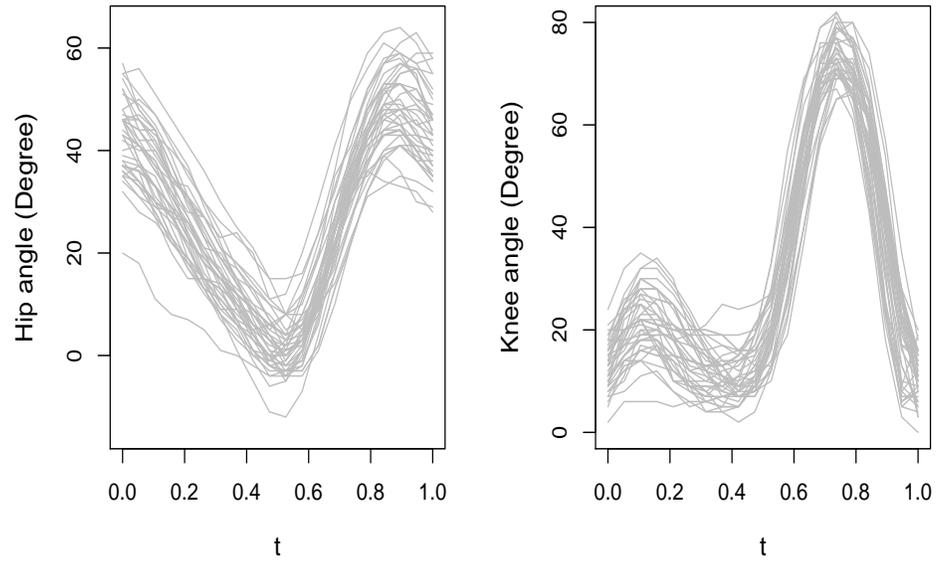}
\caption{Measurement of hip angles and knee angles in the gait study. }
\label{fig:webfig4}
\end{figure}

\begin{figure}[ht]
\centering
\scalebox{0.8}{\includegraphics{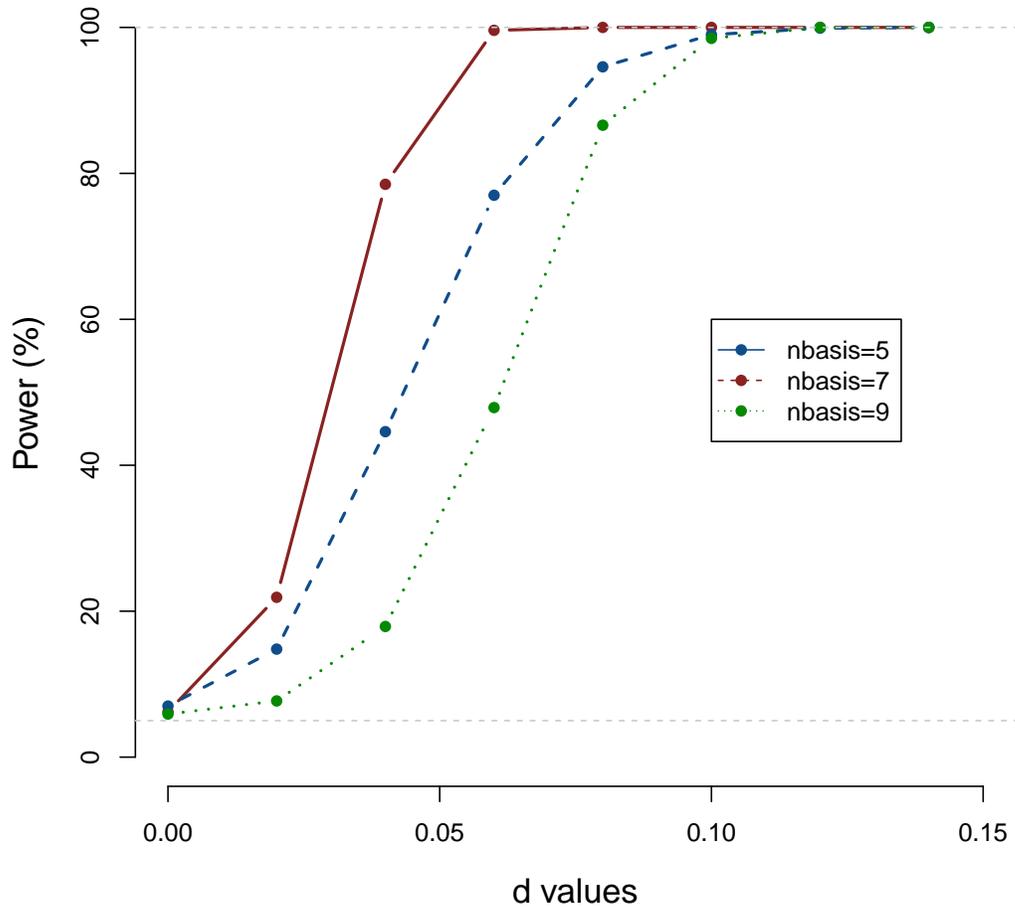}}\\
\caption{Results from simulation study mimicking gait data. Displayed is the power curve of our test. }
\label{fig:fig7}
\end{figure}

\end{document}


\title{\textbf{Supplemental Material}\\
A Score Based Test for Functional Linear Concurrent Regression}
\author{Rahul Ghosal$^{*}$and
Arnab Maity\\[4pt]
}
\date{}
\markboth%
{R.Ghosal and A.Maity}
{A Score Based Test for Functional Linear Concurrent Regression}

\maketitle
\section{Appendix S1}
{\bf  Proof of Theorem 2.2:}\\
Suppose $\tilde{\bm\theta}$ (MLE under null) and $\hat{\#\Sigma}$ are consistent estimators of $\bm\theta_0$ and $\#\Sigma$ in the sense $\tilde{\bm\theta}\overset{p}{\rightarrow}\bm\theta_0$ and
$||\hat{\#\Sigma}^{-1}- \#\Sigma^{-1}||_{2} = o_{p}(1)$ (spectral norm).
We want to show $T_S(\tilde{\bm\theta},\hat{\#\Sigma})\overset{d}{\rightarrow}T_S(\bm\theta_0,\#\Sigma)$.

We note that in the proof of Theorem 1,  $$\frac{S_{\tau_1}^2(\bm\theta_0)}{\Lambda(\bm\theta_0)}=\frac{[1/2\{tr(\frac{\^Z^{T}\^V^{-1}\^Z}{n})-(\^V^{-1/2}\*Y)^{T}\frac{\^V^{-1/2}\^Z\^Z^{T}\^V^{-1/2}}{n}(\^V^{-1/2}\*Y)\}]^{2}}{\Lambda_n(\bm\theta_0)}=\frac{S_{\tau_1,n}^2(\bm\theta_0)}{\Lambda_n(\bm\theta_0)},$$ where $S_{\tau_1,n}(\bm\theta_0)=S_{\tau_1}(\bm\theta_0)/n$ and $\^V=V(\bm\theta_0,\#\Sigma)=\#\Sigma+\tau_0^{*}\^B\^B^{T}$. Thus we have $T_S(\bm\theta_0,\#\Sigma)= \frac{S_{\tau_1,n}^2(\bm\theta_0)}{\Lambda_n(\bm\theta_0)}I(S_{\tau_1,n}(\bm\theta_0)\geq 0)$. We now show $S_{\tau_1,n}(\tilde{\bm\theta})\overset{d}{\rightarrow}S_{\tau_1,n}(\bm\theta_0)$ and $\Lambda_n(\tilde{\bm\theta})\overset{p}{\rightarrow}\Lambda_n(\bm\theta_0)$, which then by Slutsky's s theorem proves $T_S(\tilde{\bm\theta},\hat{\#\Sigma})\overset{d}{\rightarrow}T_S(\bm\theta_0,\#\Sigma)$.

$S_{\tau_1,n}(\tilde{\bm\theta})=[-1/2\{tr(\frac{\^Z^{T}\tilde{\^V}^{-1}\^Z}{n})-(\*Y)^{T}\frac{\tilde{\^V}^{-1}\^Z\^Z^{T}\tilde{\^V}^{-1}}{n}(\*Y)\}]$.
Where $\tilde{\^V}=V(\tilde{\bm\theta},\hat{\#\Sigma})=\hat{\#\Sigma}+\tilde{\tau_0}^{*}\^B\^B^{T}$. It is enough to show $||\frac{\^Z^{T}\tilde{\^V}^{-1}\^Z}{n}-\frac{\^Z^{T}\^V^{-1}\^Z}{n}||_2=o_{p}(1)$ and $||\frac{\tilde{\^V}^{-1}\^Z\^Z^{T}\tilde{\^V}^{-1}}{n}-\frac{{\^V}^{-1}\^Z\^Z^{T}{\^V}^{-1}}{n}||_2=o_{p}(1)$ to conclude
 $S_{\tau_1,n}(\tilde{\bm\theta})\overset{d}{\rightarrow}S_{\tau_1,n}(\bm\theta_0)$. Using Woodbury matrix identity, we can write  
 \begin{equation}
 \frac{\^Z^{T}\tilde{\^V}^{-1}\^Z}{n}
 =\frac{\^Z^{T}\hat{\#\Sigma}^{-1}\^Z}{n}-\tilde{\tau_0}^{*}\left(\frac{\^Z^{T}\hat{\#\Sigma}^{-1}\^B}{n}\right)\left(\frac{\^I}{n}+\frac{\tilde{\tau_0}^{*}\^B^{T}\hat{\#\Sigma}^{-1}\^B}{n}\right)^{-1}\left(\frac{\^Z^{T}\hat{\#\Sigma}^{-1}\^B}{n}\right)^T.
 \end{equation}
 We prove that each part in the r.h.s of (1) converges to the corresponding part of the right hand side, with ${\tau_0}$ and $\#\Sigma$ in place of $\tilde{\tau_0}^{*}$ and $\hat{\#\Sigma}$, and also show that each individual piece is bounded. We observe $||\frac{\^Z^{T}\hat{\#\Sigma}^{-1}\^Z}{n}-\frac{\^Z^{T}\#\Sigma^{-1}\^Z}{n}||_2 \leq||\frac{\^Z^T}{\sqrt[]{n}}||_2\hspace{1mm}||\hat{\#\Sigma}^{-1}- \#\Sigma^{-1}||_{2}\hspace{1mm}||\frac{\^Z}{\sqrt[]{n}}||_2 \leq ||\frac{\^Z^T}{\sqrt[]{n}}||_F\hspace{1mm}||\hat{\#\Sigma}^{-1}- \#\Sigma^{-1}||_{2}\hspace{1mm}||\frac{\^Z}{\sqrt[]{n}}||_F$.
 Here $||\cdot||_F$ denotes the Frobenius norm and $||\^A||_2\leq ||\^A||_F$. Note as defined in the main paper $\^Z=\^X\#\Sigma_1^{1/2}$. We then have
$||\frac{\^Z^T}{\sqrt[]{n}}||_F=||\frac{\^Z}{\sqrt[]{n}}||_F
=||\frac{\^X\#\Sigma_1^{1/2}}{\sqrt[]{n}}||_F\leq ||\#\Sigma_1^{1/2}||_F\hspace{2mm}\sqrt[]{tr(\frac{\^X^T\^X}{n})}< D$ (some $D>0$), as $||\#\Sigma_1^{1/2}||_F$ is bounded; $\#\Sigma_1^{1/2}$ being a fixed $k_1\times k_1$ matrix. Also note that $tr(\^X^{T}\^X/n)=||\^X||_{F}^2/n=\sum_{j=1}^{m}\sum_{i=1}^{n}X_{i}^2(t_j)/n\sum_{k=1}^{k_{1}}B_{k1}(t_j)^2\rightarrow\sum_{j=1}^{m}\mu_2(t_j)\sum_{k=1}^{k_{1}}B_{k1}(t_j)^2 < \infty$ by S.L.L.N, as $X_{i}$'s were assumed to be i.i.d. with finite second moment. So   
$||\frac{\^Z^{T}\hat{\#\Sigma}^{-1}\^Z}{n}-\frac{\^Z^{T}\#\Sigma^{-1}\^Z}{n}||_2=o_p(1)$, as $||\hat{\#\Sigma}^{-1}- \#\Sigma^{-1}||_{2}=o_p(1)$
 and the other terms are bounded as shown above.
 \\ Now similarly $||\frac{\^Z^{T}\hat{\#\Sigma}^{-1}\^B}{n}-\frac{\^Z^{T}\#\Sigma^{-1}\^B}{n}||_2  \leq ||\#\Sigma_1^{1/2}||_F\hspace{2mm}\sqrt[]{tr(\frac{\^X^T\^X}{n})}\hspace{2mm}||\hat{\#\Sigma}^{-1}- \#\Sigma^{-1}||_{2}
\hspace{2mm}||\#\Sigma_0^{1/2}||_F\hspace{2mm}\sqrt[]{tr(\frac{\mathcal{B}^T\mathcal{B}}{n})}=o_p(1)$. ($\^B=\mathcal{B}\#\Sigma_0^{1/2}$, $tr(\mathcal{B}^{T}\mathcal{B}/n)=||\mathcal{B}||_{F}^2/n=\sum_{j=1}^{m}\sum_{i=1}^{n}1/n\sum_{k=1}^{k_{0}}B_{k0}(t_j)^2=\sum_{j=1}^{m}\sum_{k=1}^{k_{0}}B_{k0}(t_j)^2 < \infty$). Also using sub multiplicativity of norms, $||\left(\frac{\^I}{n}+\frac{\tilde{\tau_0}^{*}\^B^{T}\hat{\#\Sigma}^{-1}\^B}{n}\right)^{-1}-\left(\frac{\^I}{n}+\frac{\tau_0\^B^{T}{\#\Sigma}^{-1}\^B}{n}\right)^{-1}||_2$\\$\leq
||\left(\frac{\^I}{n}+\frac{\tilde{\tau_0}^{*}\^B^{T}\hat{\#\Sigma}^{-1}\^B}{n}\right)^{-1}||_2\hspace{2mm}||\frac{\tilde{\tau_0}^{*}\^B^{T}\hat{\#\Sigma}^{-1}\^B}{n}-\frac{\tau_0\^B^{T}{\#\Sigma}^{-1}\^B}{n}||_2\hspace{2mm}||\left(\frac{\^I}{n}+\frac{\tau_0\^B^{T}{\#\Sigma}^{-1}\^B}{n}\right)^{-1}||_2$.\newline
The middle term $||\frac{\tilde{\tau_0}^{*}\^B^{T}\hat{\#\Sigma}^{-1}\^B}{n}-\frac{\tau_0\^B^{T}{\#\Sigma}^{-1}\^B}{n}||_2\leq|\tilde{\tau_0}^{*}-\tau_0|\hspace{1mm}||\frac{\^B^{T}\hat{\#\Sigma}^{-1}\^B}{n}||_2+|\tau_0|||\frac{\^B^{T}\hat{\#\Sigma}^{-1}\^B}{n}-\frac{\^B^{T}{\#\Sigma}^{-1}\^B}{n}||_2$\\
$\leq |\tilde{\tau_0}^{*}-\tau_0|\hspace{1mm}tr(\mathcal{B}^{T}\mathcal{B}/n)\hspace{1.5mm}||\#\Sigma_0^{1/2}||_F^{2}\hspace{1mm}
||\hat{\#\Sigma}^{-1}||_2 + |\tau_0|\hspace{1mm}tr(\mathcal{B}^{T}\mathcal{B}/n)\hspace{1.5mm}||\#\Sigma_0^{1/2}||_F^{2}\hspace{1mm}
||\hat{\#\Sigma}^{-1}- \#\Sigma^{-1}||_{2}= o_p(1)$, as
$||\hat{\#\Sigma}^{-1}- \#\Sigma^{-1}||_{2}$ and $|\tilde{\tau_0}^{*}-\tau_0|$ are $o_p(1)$ (by assumption) and rest of the terms are bounded as shown earlier. We have used the fact that $||\hat{\#\Sigma}^{-1}||_2 $ is bounded because minimum eigenvalue of $\hat{\#\Sigma}$ is $>0$. This is because $\hat{\#\Sigma}$ is block diagonal, i.e, $\hat{\#\Sigma}$=diag \{${\hat{\#\Sigma}_{m\times m},\hat{\#\Sigma}_{m\times m},\ldots,\hat{\#\Sigma}_{m\times m}}$\} and $\hat{\#\Sigma}(s,t)=\sum_{k=1}^{K}\hat{\lambda_k}\hat{\phi_k}(s)\hat{\phi_k}(t) + \hat{\sigma^2} I(s=t)$. All the inequalities follow using application of triangle inequality and sub multiplicativity of norms. For showing $||\left(\frac{\^I}{n}+\frac{\tilde{\tau_0}^{*}\^B^{T}\hat{\#\Sigma}^{-1}\^B}{n}\right)^{-1}||_2$ and $||\left(\frac{\^I}{n}+\frac{\tau_0\^B^{T}{\#\Sigma}^{-1}\^B}{n}\right)^{-1}||_2$ are bounded, it suffices to show minimum eigenvalues of $\frac{\^I}{n}+\frac{\tilde{\tau_0}^{*}\^B^{T}\hat{\#\Sigma}^{-1}\^B}{n}$ and 
$\frac{\^I}{n}+\frac{\tau_0\^B^{T}{\#\Sigma}^{-1}\^B}{n}$ are positive.
Now $\frac{\^I}{n}+\frac{\tilde{\tau_0}^{*}\^B^{T}\hat{\#\Sigma}^{-1}\^B}{n}=\frac{\^I}{n}+\frac{n\tilde{\tau_0}^{*}{\^C_{0}}^{T}_{k_0\times m}\hat{\#\Sigma}_{m\times m}^{-1}{\^C_{0}}_{m\times k_0}}{n}=\frac{\^I}{n}+\tilde{\tau_0}^{*}{\^C_{0}}^{T}_{k_0\times m}\hat{\#\Sigma}_{m\times m}^{-1}{\^C_{0}}_{m\times k_0}$. Here  
${\^C_{0}}={\^B_{0}}\#\Sigma_0^{1/2}$ is full column rank,
so ${\^C_{0}}^{T}_{k_0\times m}\hat{\#\Sigma}_{m\times m}^{-1}{\^C_{0}}_{m\times k_0}$ is p.d matrix (not depending on n) and therefore minimum eigenvalue of $\frac{\^I}{n}+\frac{\tilde{\tau_0}^{*}\^B^{T}\hat{\#\Sigma}^{-1}\^B}{n}$  is positive. In similar way it can be shown, minimum eigenvalues of $\frac{\^I}{n}+\frac{\tau_0\^B^{T}{\#\Sigma}^{-1}\^B}{n}$ are positive. This completes the proof that $||\left(\frac{\^I}{n}+\frac{\tilde{\tau_0}^{*}\^B^{T}\hat{\#\Sigma}^{-1}\^B}{n}\right)^{-1}-\left(\frac{\^I}{n}+\frac{\tau_0\^B^{T}{\#\Sigma}^{-1}\^B}{n}\right)^{-1}||_2=o_p(1)$.

So we have shown all the terms in r.h.s of (1) converge to the corresponding counterparts with true values $\bm\theta_0$ and $\#\Sigma$. Note that all the individual terms are bounded, i.e, $||\frac{\^Z^{T}\hat{\#\Sigma}^{-1}\^Z}{n}||_2$, $||(\frac{\^Z^{T}\hat{\#\Sigma}^{-1}\^B}{n})||_2$, $||\left(\frac{\^I}{n}+\frac{\tilde{\tau_0}^{*}\^B^{T}\hat{\#\Sigma}^{-1}\^B}{n}\right)^{-1}||_2$ are bounded with the bound not depending on $n$. These can be shown exactly in the same way using sub multiplicativity of spectral norms, relation between spectral and Frobenius norm and the bounds established during the proof of the convergence part.
So by repeated use of triangle inequality from (1) we have $||\frac{\^Z^{T}\tilde{\^V}^{-1}\^Z}{n}-\frac{\^Z^{T}\^V^{-1}\^Z}{n}||_2=o_{p}(1)$.

Next to complete the proof we have to show $||\frac{\tilde{\^V}^{-1}\^Z\^Z^{T}\tilde{\^V}^{-1}}{n}-\frac{{\^V}^{-1}\^Z\^Z^{T}{\^V}^{-1}}{n}||_2=o_{p}(1)$ and this is again done using similar kind of techniques.
Let us denote $\frac{{\^V}^{-1}\^Z}{\sqrt[]{n}}=\^A$, $\frac{\tilde{\^V}^{-1}\^Z}{\sqrt[]{n}}=\tilde{\^A}$. Then we are trying to show $||\tilde{\^A}\tilde{\^A}^T-\^A\^A^T||_2=o_{p}(1)$. Now using triangle and norm inequalities we have $||\tilde{\^A}\tilde{\^A}^T-\^A\^A^T||_2\leq||\tilde{\^A}||_2||\tilde{\^A}^T-\^A^T||_2 +||\tilde{\^A}^T-\^A^T||_2||{\^A}||_2 $. It suffices to show that $||\tilde{\^A}||_2,||{\^A}||_2$ are bounded and $||\tilde{\^A}^T-\^A^T||_2=o_p(1)$. Note $||{\^A}||_2=||{\^A}^T||_2=||\frac{\^Z^{T}\tilde{\^V}^{-1}}{\sqrt[]{n}}||_2$ and thus we have 
\begin{eqnarray}
 \frac{\^Z^{T}\tilde{\^V}^{-1}}{\sqrt[]{n}}
 =\frac{\^Z^{T}\hat{\#\Sigma}^{-1}}{\sqrt[]{n}}-\tilde{\tau_0}^{*}(\frac{\^Z^{T}\hat{\#\Sigma}^{-1}\^B}{n})\left(\frac{\^I}{n}+\frac{\tilde{\tau_0}^{*}\^B^{T}\hat{\#\Sigma}^{-1}\^B}{n}\right)^{-1}(\frac{\^B^{T}\hat{\#\Sigma}^{-1}}{\sqrt[]{n}}).
 \end{eqnarray}
Now as in (1), it can be shown similarly that all the terms in r.h.s of (2) are bounded and each term converges to
the corresponding counterpart with ${\tau_0}$ and $\#\Sigma$ in place of $\tilde{\tau_0}^{*}$ and $\hat{\#\Sigma}$. This would then imply $||\tilde{\^A}||_2,||{\^A}||_2$ are bounded and $||\tilde{\^A}^T-\^A^T||_2=o_p(1)$. So we have $||\frac{\tilde{\^V}^{-1}\^Z\^Z^{T}\tilde{\^V}^{-1}}{n}-\frac{{\^V}^{-1}\^Z\^Z^{T}{\^V}^{-1}}{n}||_2=o_{p}(1)$ and along with $||\frac{\^Z^{T}\tilde{\^V}^{-1}\^Z}{n}-\frac{\^Z^{T}\^V^{-1}\^Z}{n}||_2=o_{p}(1)$, this proves $S_{\tau_1,n}(\tilde{\bm\theta})\overset{d}{\rightarrow}S_{\tau_1,n}(\bm\theta_0)$. The proof $\Lambda_n(\tilde{\bm\theta})\overset{p}{\rightarrow}\Lambda_n(\bm\theta_0)$ is again based on using similar set of techniques applied so far. We note $\Lambda_n(\bm\theta_0)=\Lambda(\bm\theta_0)/n^2=\frac{1}{2}tr\{ (\^Z^{T}V^{-1}\^Z/n)^2\}$- $\frac{\left[\frac{1}{2}tr\{(\^B^{T}\^V^{-1}\^Z/n)(\^B^{T}\^V^{-1}\^Z/n)^T\}\right]^2}{\frac{1}{2}tr\{(\^B^{T}\^V^{-1}\^B/n)^2)\}}$. It was already shown that  $||\frac{\^Z^{T}\tilde{V}^{-1}\^Z}{n}-\frac{\^Z^{T}V^{-1}\^Z}{n}||_2=o_{p}(1)$, similarly it can be shown $||\frac{\^B^{T}\tilde{\^V}^{-1}\^Z}{n}-\frac{\^B^{T}\^V^{-1}\^Z}{n}||_2=o_{p}(1)$ and $||\frac{\^B^{T}\tilde{\^V}^{-1}\^B}{n}-\frac{\^B^{T}\^V^{-1}\^B}{n}||_2=o_{p}(1)$; which along with the facts, eigenvalues of square of a matrix are squares of it's eigenvalues, and trace is the sum of eigenvalues, proves $\Lambda_n(\tilde{\bm\theta})\overset{p}{\rightarrow}\Lambda_n(\bm\theta_0)$. Therefore by Slutsky's theorem $T_S(\tilde{\bm\theta},\hat{\#\Sigma})\overset{d}{\rightarrow}T_S(\bm\theta_0,\#\Sigma)\overset{d}{=} (1/2)^2\frac{\left(\sum_{\ell=1}^{k_{1}}\lambda_\ell x_{\ell}^{2}-\sum_{\ell=1}^{k_{1}}\lambda_\ell\right)^2}{\Lambda_n(\bm\theta_0)}I(\sum_{\ell=1}^{k_{1}}\lambda_\ell x_{\ell}^{2}\geq \sum_{\ell=1}^{k_{1}}\lambda_\ell)$, where $x_{\ell}\stackrel{iid}{\sim}\mathcal{N}(0,1)$ and $\lambda_\ell$ are eigenvalues of $\frac{\^Z^{T}\^V(\bm\theta_0,\#\Sigma)^{-1}\^Z}{n}$. 

As $||\frac{\^Z^{T}\tilde{\^V}^{-1}\^Z}{n}-\frac{\^Z^{T}\^V^{-1}\^Z}{n}||_2=o_{p}(1)$ and $\Lambda_n(\tilde{\bm\theta})\overset{p}{\rightarrow}\Lambda_n(\bm\theta_0)$, we approximate the null distribution using $
(1/2)^2\frac{\left(\sum_{\ell=1}^{k_{1}}\tilde{\lambda_\ell}x_{\ell}^{2}-\sum_{\ell=1}^{k_{1}}\tilde{\lambda_\ell}\right)^2}{\Lambda_n(\tilde{\bm\theta})}I(\sum_{\ell=1}^{k_{1}}\tilde{\lambda_\ell}x_{\ell}^{2}\geq \sum_{\ell=1}^{k_{1}}\tilde{\lambda_\ell})
$, where $\tilde{\lambda_\ell}$ are eigenvalues of $\frac{\^Z^{T}\^V(\tilde{\bm\theta},\hat{\#\Sigma})^{-1}\^Z}{n}$ and $x_{\ell}\stackrel{iid}{\sim}\mathcal{N}(0,1)$ for $\ell=1,2,\ldots,k_1$.\par
\newpage
\section{Appendix S2: Asymptotic distribution of p-value under null}
\textbf{Case 1: $a=0$}\\
$P(T_S(\tilde{\bm\theta})\geq a){\rightarrow} P(T_S(\bm\theta_0)\geq a)$
\vspace*{-2 mm}
\begin{align*}
&= P(T_S(\bm\theta_0)\geq 0|S_{\tau_1}(\bm\theta_0)<0)(1-\alpha) + P(T_S(\bm\theta_0)\geq 0|S_{\tau_1}(\bm\theta_0)\geq 0)\alpha\\
&=P(\frac{S_{\tau_1}^2(\bm\theta_0)}{\Lambda(\bm\theta_0)}I(S_{\tau_1}(\bm\theta_0)\geq0)\geq 0|S_{\tau_1}(\bm\theta_0)<0)(1-\alpha)\\
& + P(\frac{S_{\tau_1}^2(\bm\theta_0)}{\Lambda(\bm\theta_0)}I(S_{\tau_1}(\bm\theta_0)\geq 0)\geq 0|S_{\tau_1}(\bm\theta_0)\geq0)\alpha\\
&=1\times(1-\alpha)+1\times \alpha =1.
\end{align*}
where $\alpha=P_{H_{0}}(S_{\tau_1}(\theta_{0})\geq 0)=P_{H_{0}}(S_{\tau_1,n}(\theta_{0})\geq 0)=P(\sum_{\ell=1}^{k_{1}}\lambda_\ell x_\ell^{2}\geq\sum_{\ell=1}^{k_{1}}\lambda_\ell)$.\\
\textbf{Case 2: $a > 0$}\\
$P(T_S(\tilde{\bm\theta})\geq a){\rightarrow} P(T_S(\bm\theta_0)\geq a)$
\vspace*{-2 mm}
\begin{align*}
&= P(T_S(\bm\theta_0)\geq a|S_{\tau_1}(\bm\theta_0)<0)(1-\alpha) + P(T_S(\bm\theta_0)\geq a|S_{\tau_1}(\bm\theta_0)\geq 0)\alpha\\
&=P(\frac{S_{\tau_1}^2(\bm\theta_0)}{\Lambda(\bm\theta_0)}I(S_{\tau_1}(\bm\theta_0)\geq 0)\geq a|S_{\tau_1}(\bm\theta_0)<0)(1-\alpha)\\
&+ P(\frac{S_{\tau_1}^2(\bm\theta_0)}{\Lambda(\bm\theta_0)}I(S_{\tau_1}(\bm\theta_0)>0)\geq a|S_{\tau_1}(\bm\theta_0)\geq 0)\alpha\\
&=0\times (1-\alpha)+\alpha\times P(\frac{S_{\tau_1}^2(\bm\theta_0)}{\Lambda(\bm\theta_0)}I(S_{\tau_1}(\bm\theta_0)\geq 0)\geq a|S_{\tau_1}(\bm\theta_0)\geq 0) \\
&=\alpha\times P(\frac{S_{\tau_1}^2(\bm\theta_0)}{\Lambda(\bm\theta_0)}I(S_{\tau_1}(\bm\theta_0)\geq 0)\geq a|S_{\tau_1}(\bm\theta_0)\geq 0)\\
&=\alpha\times \frac{P(\frac{S_{\tau_1}^2(\bm\theta_0)}{\Lambda(\bm\theta_0)}I(S_{\tau_1}(\bm\theta_0)\geq0)\geq a \cap S_{\tau_1}(\bm\theta_0)\geq 0)}{P(S_{\tau_1}(\bm\theta_0)\geq 0)}\\
&=\alpha\times \frac{P(\frac{S_{\tau_1}^2(\bm\theta_0)}{\Lambda(\bm\theta_0)}I(S_{\tau_1}(\bm\theta_0)\geq 0)\geq a \cap S_{\tau_1}(\bm\theta_0)\geq 0)}{\alpha}\\
&=P(\frac{S_{\tau_1}^2(\bm\theta_0)}{\Lambda(\bm\theta_0)}I(S_{\tau_1}(\bm\theta_0)\geq 0)\geq a \cap S_{\tau_1}(\bm\theta_0)\geq 0)\\
&=P(S_{\tau_1}^2(\bm\theta_0)\geq a\Lambda(\bm\theta_0)\hspace{.2 cm} \cap S_{\tau_1}(\bm\theta_0)\geq 0)
=P(S_{\tau_1,n}(\bm\theta_0))\geq \sqrt{a\Lambda_n(\bm\theta_0)})\\
&=P\left(1/2(\sum_{\ell=1}^{k_{1}}\lambda_{\ell}x_\ell^{2}-\sum_{\ell=1}^{k_{1}}\lambda_\ell)\geq \sqrt{a\Lambda_n(\bm\theta_0)}\right)\\
&=P\left(\sum_{\ell=1}^{k_{1}}\lambda_\ell x_{\ell}^{2}\geq \sum_{l=\ell}^{k_{1}}\lambda_\ell+2\sqrt{\Lambda_n(\bm\theta_0)}\sqrt{a}\right)
\leq P\left(\sum_{\ell=1}^{k_{1}}\lambda_\ell x_{\ell}^{2}\geq \sum_{\ell=1}^{k_{1}}\lambda_\ell\right)=\alpha.
\end{align*}
From this intuitively it is clear that under null p-value asymptotically takes value from a mixture distribution of uniform$(0,\alpha)$ and a degenerate distribution at one. Next we prove this fact. Again all the calculations are done assuming null is true.

Let $P_v$ stand for p-value. We can define the p-value as $P_v(c)=P(T_s > T_{obs}|T_{obs}=c)$.\\
Then from the previous discussion it follows that
$$P(P_v(T_{obs}) = 1)=P(T_{obs}=0)=P(S_{\tau_1}(\tilde{\bm\theta})\leq0){\rightarrow}P(S_{\tau_1}(\bm\theta_0)\leq 0)=1-\alpha.$$
Recall that $T_s(\bm\theta_0)=\frac{S_{\tau_1}^2(\bm\theta_0)}{\Lambda(\bm\theta_0)}I(S_{\tau_1}(\bm\theta_0)\geq0)$. So for $b>0$, $P(T_s\geq b)=\alpha\overline{G(b)}$, where
$G(b)$ is distribution function of the continuous part of $T_S$ namely $\frac{S_{\tau_1}^2(\bm\theta_0)}{\Lambda(\bm\theta_0)}$ and $\overline{G(b)}=1-G(b)$.

Hence we have for $0\leq d \leq \alpha<1$, 
\begin{eqnarray*}
P(p_v(T_{obs})\leq d) &=& P(P(T_S\geq T_{obs})\leq d)\\
&=& P(P(T_S\geq T_{obs})\leq d|T_{obs}>0)P(T_{obs}>0)\\
&{\rightarrow}& P(P(T_S\geq T_{obs})\leq d|T_{obs}>0)\alpha.
\end{eqnarray*}
The last convergence follows since $P(T_{obs}>0)=P(S_{obs}>0) {\rightarrow} P(S_{\tau_1}(\bm\theta_0)>0)=P(S_{\tau_1}(\bm\theta_0)\geq 0)=\alpha$, asymptotically under null, as $\tilde{\bm\theta}$ is a consistent estimator of ${\bm\theta_0}$ and $S_{\tau_1}(\cdot)$ is a continuous random variable . Thus it is easy to see that
\begin{eqnarray*}
P(p_v(T_{obs})\leq d) & \rightarrow &P(\alpha\overline{G(T_{obs})}\leq d)\alpha\\
&=& P(\overline{G(T_{obs})}\leq d/\alpha)\alpha\\
&=& (d/\alpha)\alpha = d.
\end{eqnarray*}
The last line follows because  $\left(\overline{G(T_{obs})}\right) \sim U(0,1)$, as $G$ is a continuous distribution function. 
So under null p-value is asymptotically an $\alpha$:$(1-\alpha)$ mixture of uniform$(0,\alpha)$ and degenerate distribution at one.

\newpage
\section{Supporting Figures}
Figure S1 displays the power curve for a simulation study for dense data as in Scenario A in the paper, with sample size $n=39$, and $m=20$. Figure S2 shows the effect of number of basis on the power of the  proposed test for simulation Scenario A, dense design, with sample size $n=300$. Figure S3 displays the same for Scenario Bt. Figure S4 plots the observed calcium absorption and calcium intake of the patients along their ages in the calcium absorption study.  

\begin{figure}[ht]
\centering
\includegraphics[width=.9\linewidth , height=.85\linewidth]{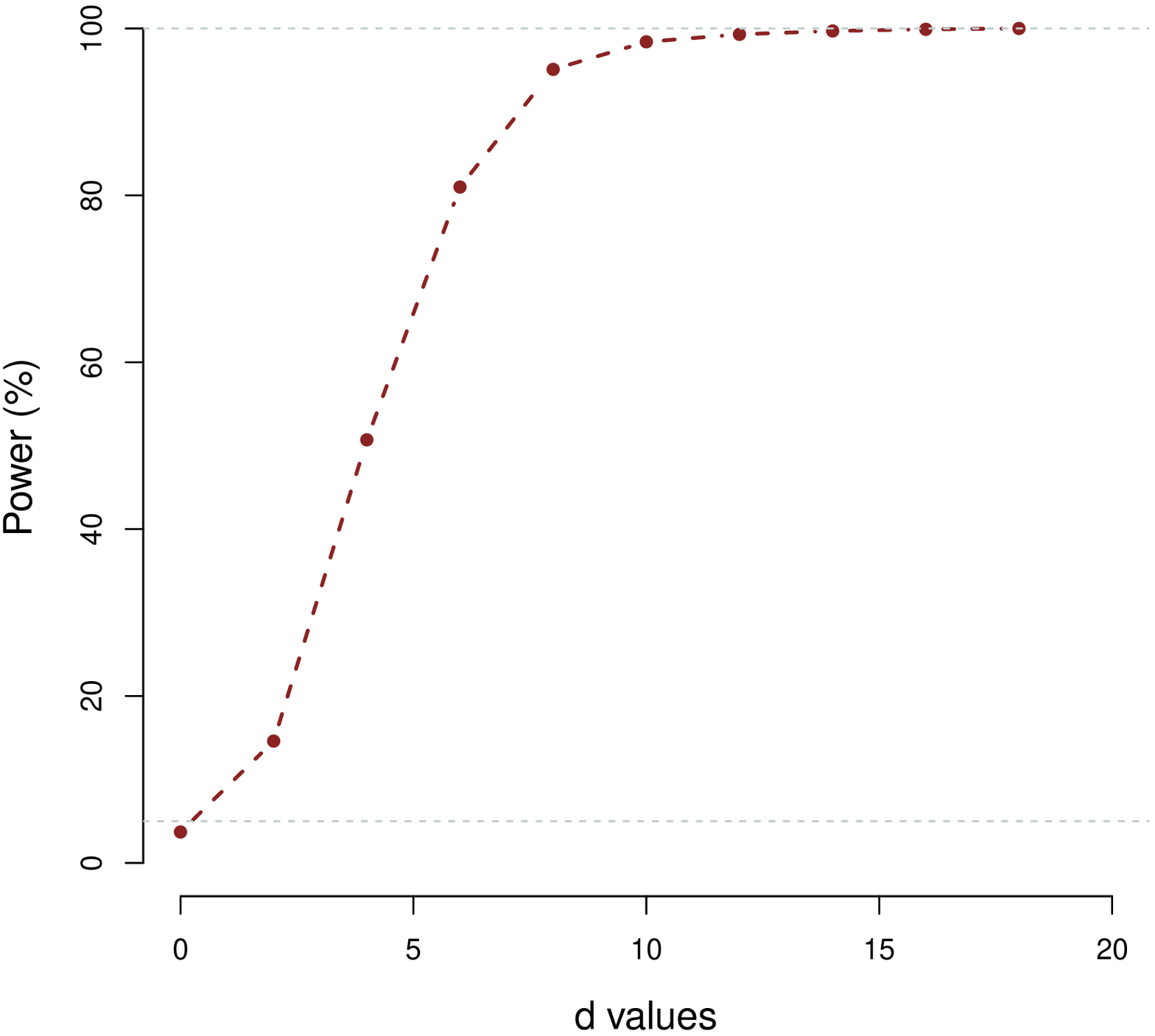}
\caption{Power curve for simulation Scenario A, n=39, m=20.}
\label{fig:webfig2}
\end{figure}

\begin{figure}[ht]
\centering
\includegraphics[width=.9\linewidth , height=.85\linewidth]{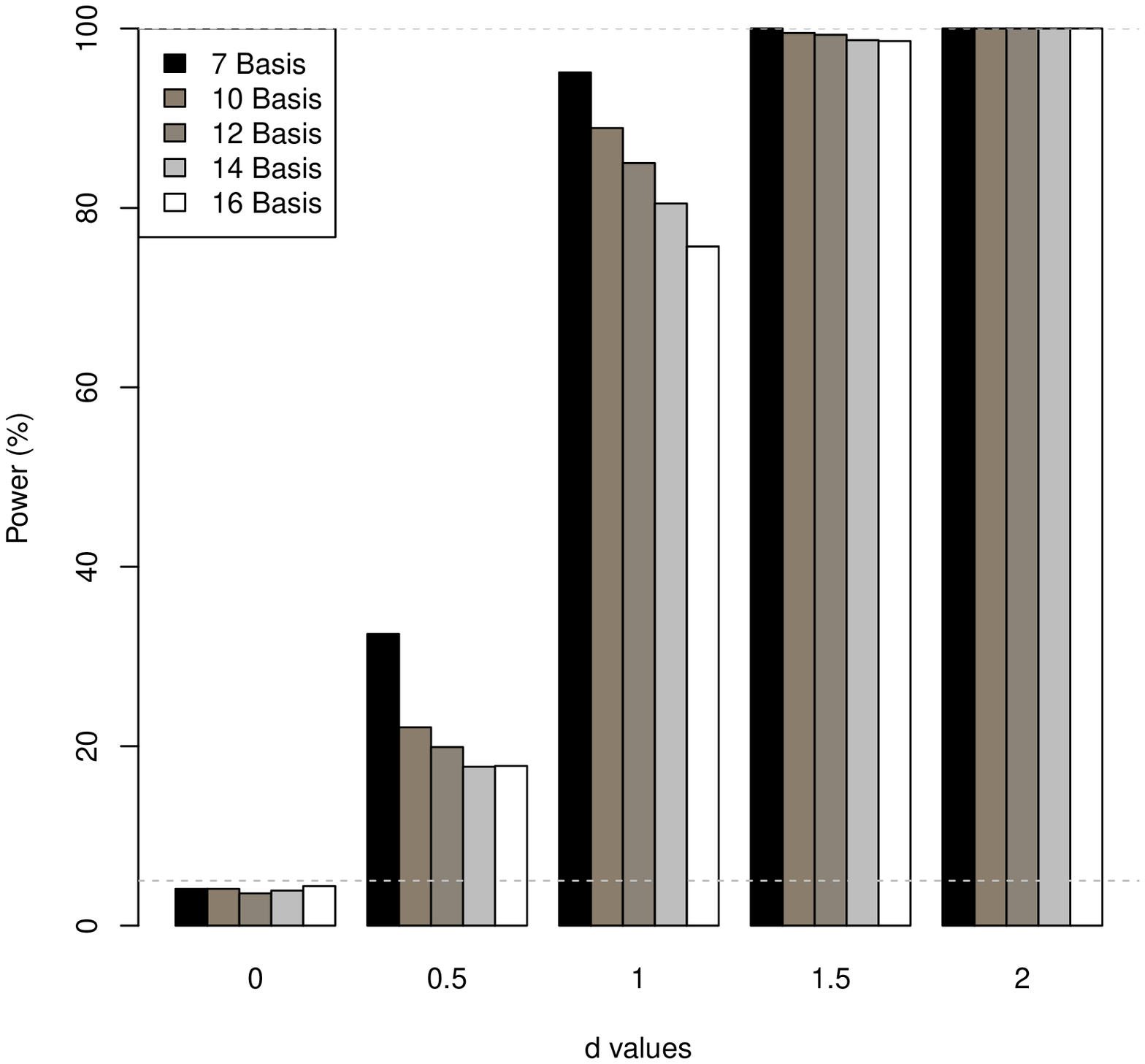}
\caption{Effect of number of basis on the power of the test, simulation Scenario A, n=300.}
\label{fig:webfig3}
\end{figure}

\begin{figure}[ht]
\centering
\includegraphics[width=.9\linewidth , height=.85\linewidth]{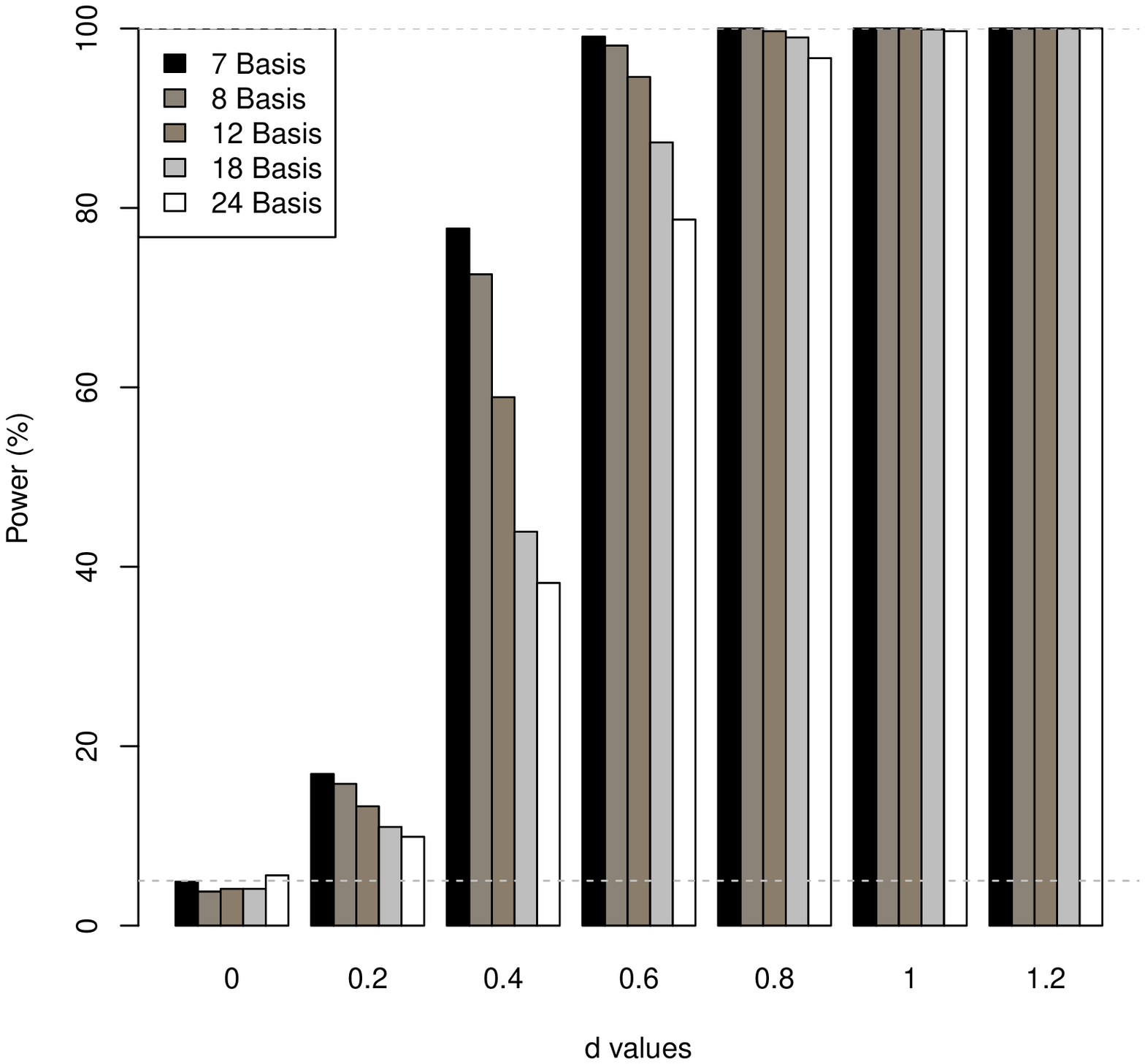}
\caption{Effect of number of basis on the power of the test, simulation Scenario B, n=100.}
\label{fig:webfig3}
\end{figure}
   

\begin{figure}[ht]
\centering
\includegraphics[width=.9\linewidth , height=.65\linewidth]{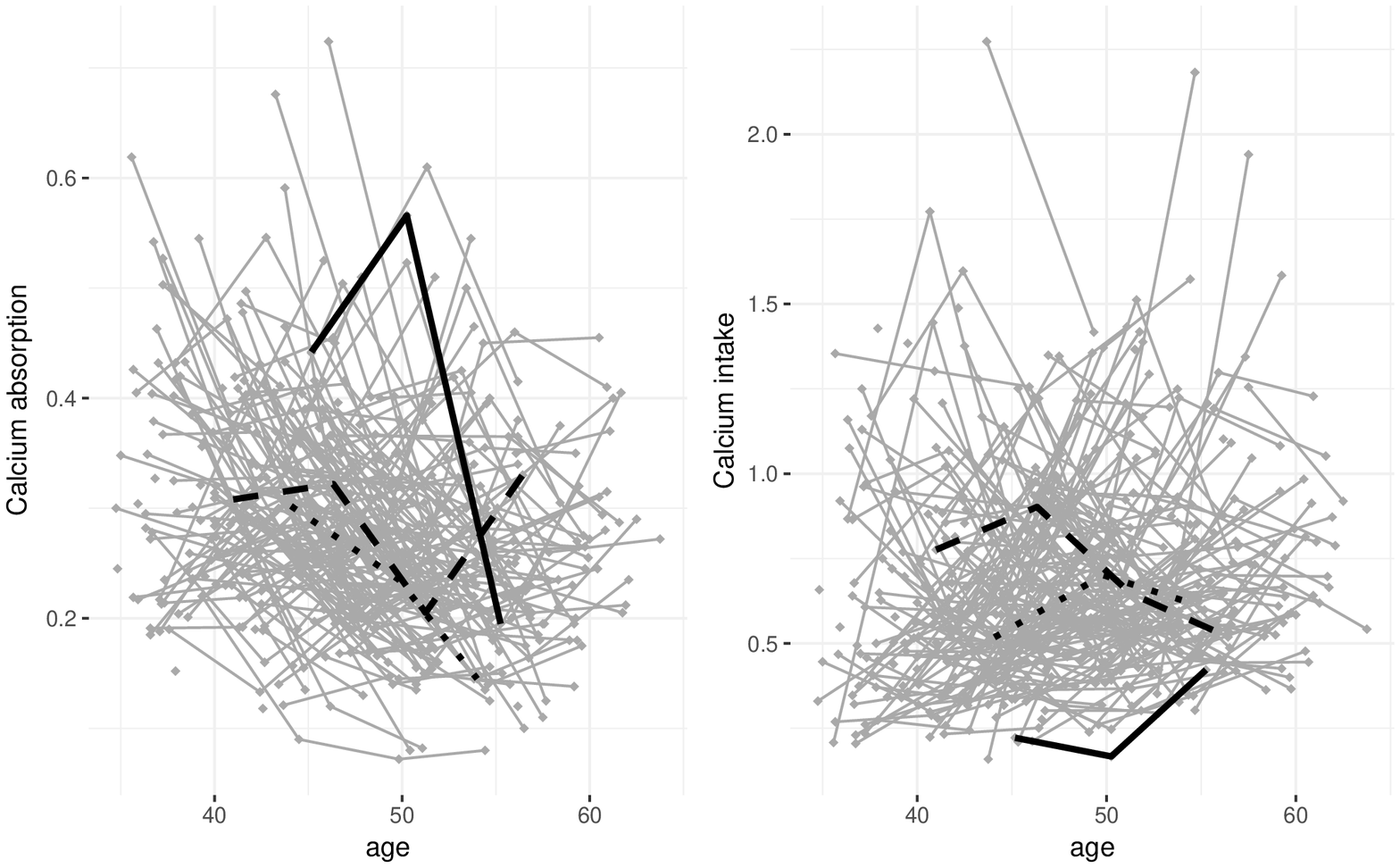}
\caption{Observed calcium absorption and calcium intake in the calcium absorption study.}
\label{fig:webfig5}
\end{figure}